\documentclass[aps,preprintnumbers,onecolumn,amsmath,amssymb,floatfix,pra]{revtex4} 
\pdfoutput=1
\usepackage{graphicx}
\usepackage{color}

\usepackage{ifpdf}

\ifpdf
\usepackage{epstopdf}
\usepackage[pdftex]{hyperref}
\else
\usepackage[hypertex,ps2pdf,dvips,colorlinks,urlcolor=blue,citecolor=blue]{hyperref}
\fi

\newcommand\be{\begin{equation}}
\newcommand\bea{\begin{eqnarray}}
\newcommand\bes{\begin{subequations}}
\newcommand\esu{\end{subequations}}
\newcommand\ee{\end{equation}}
\newcommand\eea{\end{eqnarray}}

\newcommand\p             {\partial}

\newcommand\doi[2]        {\href{http://dx.doi.org/#1}{#2}}

\begin{document}

\title{Interaction quench in a trapped one-dimensional Bose gas}

\author{Paolo P. Mazza$^1$,
Mario Collura$^1$, 
M\'arton Kormos$^{2}$, 
and Pasquale Calabrese$^1$
}

\affiliation{
$^1$Dipartimento di Fisica dell'Universit\`a di Pisa and INFN, 56127 Pisa, Italy\\
$^2$ MTA-BME ÒMomentumÓ Statistical Field Theory Research Group, 1111 Budapest, Budafoki \'ut 8, Hungary 
}

\date{\today}

\begin{abstract}

We study the non-equilibrium quench dynamics from free to hard-core one-dimensional bosons in the presence 
of a hard-wall confining potential. 
We characterise the density profile and the two-point fermionic correlation function in the stationary state as 
well as their full time evolution. 
We find that for long times the system relaxes to a uniform density profile, but the correlation function keeps memory 
of the initial state with a stationary algebraic long-distance decay as opposite to the exponential behaviour found for the 
same quench in the periodic setup. 
We also compute the stationary bosonic two-point correlator which turns out to decay exponentially for large 
distances. 
We show that a two-step mechanism governs the time evolution: a quick approach to an almost stationary value 
is followed by a slow algebraic relaxation to the true stationary state.

\end{abstract}


\maketitle

\section{Introduction}

The non-equilibrium dynamics of isolated quantum systems is currently in a golden age mainly because of the 
recent experiments on trapped ultra-cold atomic gases
\cite{uc,kww-06,tetal-11,cetal-12,getal-11,shr-12,rsb-13} which allowed for the realization and the experimental study of  
the unitary non-equilibrium evolution over long time scales.  
The non-equilibrium situation which attracted most of the theorists' attention is the so called interaction quench, in which 
a system evolves unitarily from an initial state which is the ground-state of a translationally invariant Hamiltonian 
differing from the one governing the evolution  by an experimentally tunable interaction parameter \cite{revq}.

One of the most interesting findings of both theoretical and experimental investigations is the different behaviour displayed 
by generic and integrable systems,
with the latter keeping memory of the initial state also for infinite time \cite{kww-06,gg,rs-12,ce-13} while the 
former locally relaxing to a standard 
Gibbs distribution in which only the initial energy determines the (local) stationary state as an effective temperature 
\cite{nonint,rdo-08,tvar}. 
However, most of the previous studies lack the direct connection to the experiments in which the atoms are trapped 
by some external potential, a situation that for a truly interacting model is very difficult  (if not impossible) to tackle analytically
in an exact way.
For this reason,  we consider here one of the simplest instances of an interaction quench in the presence of a 
simple confining potential. Despite of this double level of simplicity, we shall see that 
the calculations are non-trivial and that very interesting effects appear in the quench dynamics.   

We consider a one-dimensional Bose gas with Hamiltonian 
\be
\hat H = \int_{0}^{L} dx \Big[\partial_x \hat\phi^{\dagger} (x) \partial_x \hat\phi (x) 
+ c\, \hat\phi^{\dagger} (x) \hat\phi^{\dagger} (x) \hat\phi (x) \hat\phi(x)  + V(x) \hat\phi^{\dagger} (x) \hat\phi (x)  \Big].
\label{HLL}
\ee
Here $\hat\phi(x)$ is a boson field satisfying canonical commutation relations
$[\hat\phi(x),\hat\phi^{\dag}(y)]=\delta(x-y)$, $c$ is the two-body coupling constant, $V(x)$ is the confining potential and we 
set $\hbar=2m=1$. 
In the absence of the external potential (i.e. for $V(x)=0$) the Hamiltonian reduces to the celebrated 
Lieb--Liniger Hamiltonian \cite{LiebPR130} which is integrable and exactly solvable by Bethe ansatz 
for any value of the interaction strength $c$. 
Global quenches of the coupling constant $c$ have already been studied in several
papers \cite{grd-10,fle-10,mc-12,ksc-13,nm-13,kcc14,nwbc-13,ds-13,ckc-14,b-14,cd-14},
as well as other interesting quench dynamics \cite{mg-05,cro,ck-12,a-12,v-12,csc13,m-13,gn-14b}.
However, in the presence of an external potential $V(x)$, the Hamiltonian (\ref{HLL}) is not integrable 
for arbitrary values of the coupling constant $c$. 
There are only two special points in which the model is still exactly solvable for arbitrary $V(x)$ which 
correspond to free bosons ($c=0$) and impenetrable bosons ($c=+\infty$).
Indeed, for periodic boundary conditions (PBC), the quench from $c=0$ to $c=\infty$ has already been studied 
in Ref. \cite{kcc14} and despite of the simplicity of the initial and final Hamiltonian, the non-equilibrium dynamics 
turned out to be extremely rich (e.g. breaking of Wick's theorem for finite time) because the initial and final modes
are not linearly related. 

Consequently, it is very interesting to obtain analytical results for a quench from noninteracting to strongly 
interacting bosons for a trapped gas. 
The most natural choice for the confining potential would be a harmonic one (i.e. $V(x)\propto x^2$) which is the 
most commonly used in experiments.
However, while it is possible but cumbersome to perform analytic calculations with a harmonic trap,
this does not represent the easiest choice to introduce and understand the new effects engendered by the trap.   
The simplest confining potential which gives rise to most of the relevant trapping effects is surely  
the hard-wall potential which forces the many-body wave function to vanish
outside a given interval of length $L$ (this can be seen as a power-law confining trap $V(x)\propto |2x/L|^\alpha$ 
in the limit of large exponent $\alpha \gg 1$).
For all these reasons, we limit ourselves to consider the quench in a hard-wall trap whose 
main nontrivial aspect is that the initial state, i.e. the Bose-Einstein condensate (BEC) in the trap, breaks translational invariance. 
As we shall see, this leads to a number of unexpected results which we briefly anticipate now. 
First, although the initial state is highly inhomogeneous and the Hamiltonian governing the dynamics breaks translational invariance, 
in the large-time limit the density becomes homogeneous (sufficiently far from the boundaries).
However, the stationary fermionic two-point function is very different from the periodic case. 
Indeed, while for PBC it decays exponentially for large distances \cite{kcc14}, in the presence of the hard wall
trap the decay is only algebraic for points deeply in the bulk of the system.
This is a very unexpected result because it physically means that the system keeps memory of the
inhomogeneity of the initial state, even if the density becomes constant. 
Furthermore, this is  different from what would have happened if the system had thermalised because, 
at finite temperature, the boundary conditions can only affect a small region close to the boundaries and not the bulk of the system.

The paper is organised as follows. 
In Section \ref{sec2} we introduce the model under investigation and the quench protocol; 
in particular, we focus our attention on the {\it nonlinear} mapping between pre- and post-quench 
field operators and we stress the nontrivial aspects introduced by the confining potential.
In Section \ref{sec3} we analyse the stationary behaviour of the density and of the two-point function
which could be equivalently described in terms of a generalised Gibbs ensemble (GGE).
Section \ref{sec4} is devoted to the analytical evaluation of the full time-dependence of 
the particle density and of the two-point fermionic correlators.  
We also exactly characterise how the stationary values are approached for infinite time.
Finally, in Section \ref{sec5} we draw our conclusions.

\section{Model and Quench}\label{sec2}

We consider a one-dimensional Bose gas described by the Hamiltonian (\ref{HLL})
with a hard-wall confining potential on the interval $[0,L]$, i.e. 
the potential forces the many-body wave function to vanish at the boundaries $x=0,L$.
It is worth mentioning that, in the case of hard-wall trap, the Lieb-Liniger model is integrable for 
arbitrary values of the coupling constant $c$ \cite{hwint}.
However, in what follows, we limit to consider the out-of-equilibrium unitary dynamics generated by an interaction quench
of the coupling constant $c$, from noninteracting bosons ($c=0$) to hard-core bosons ($c=\infty$) in the
presence of the hard-wall boundaries (HBC) at all times.

\subsection{The initial setup}
The many-body system is initially prepared in the $N$-particle ground state of the free-boson Hamiltonian,
i.e. Eq. (\ref{HLL}) with $c=0$.
Since the Hamiltonian is quadratic, it can be diagonalised in terms of the modes 
\be\label{xi_q}
\hat \xi_{q} =\int_{0}^{L} dx \, \varphi^{*}_q(x) \, \hat\phi(x), 
\quad \hat \xi^{\dag}_{q} = \int_{0}^{L} dx \,  \varphi_q(x) \, \hat\phi^{\dag}(x), 
\ee 
where the normalised one-particle eigenfunctions
\be\label{eigenfunctions}
\varphi_q(x) =\sqrt{\frac{2}{L}}\sin(q\pi x /L), \quad q=1,2,\dots,
\ee
are the solutions of the one-particle eigenvalue problem 
\be
\left\{\begin{array}{c}\partial^2_x \varphi_{q}(x) = \epsilon_q \varphi_q(x), \\
\\
 \varphi_q(0) = \varphi_q(L) = 0, \end{array}\right. 
\ee
with $\epsilon_q = (q\pi/L)^2$.
Indeed, by using the inverse of the  transformation (\ref{xi_q}),
we can rewrite the initial Hamiltonian $\hat H_0$ in the diagonal form
\be\label{H_0_diag}
\hat H_0 = \sum_{q=1}^\infty \epsilon_q \, \hat \xi^{\dag}_q \hat\xi_q.
\ee
As usual for a BEC, the many-body ground state is prepared by filling the lowest energy level ($q=1$) with $N$ particles:
\be\label{GS}
 |\psi_0(N)\rangle=\frac1{\sqrt{N!}}\hat\xi^N_{1}|0\rangle,
 \ee
 where $|0\rangle$ is the pre-quench vacuum state  characterised by $\hat\xi_q |0\rangle = 0$.
 The initial two-point bosonic correlation function 
 $\langle \psi_0(N) | \hat\phi^{\dag}(x)\hat\phi(y)| \psi_0(N)\rangle$
 can be evaluated, for each finite value of $N$ and $L$, 
 by exploiting the canonical bosonic algebra $[\xi_{p},\xi^{\dag}_{q}]=\delta_{p,q}$ and 
 $\langle \psi_0(N) | \hat\xi^{\dag}_p \hat\xi_q| \psi_0(N)\rangle = N \delta_{p,1}\delta_{q,1}$, from which one obtains 
 \be
 \langle \psi_0(N) | \hat\phi^{\dag}(x)\hat\phi(y)| \psi_0(N)\rangle = 2 n \sin(\pi x /L) \sin(\pi y /L),\qquad n\equiv N/L.
 \ee
In particular, the initial particle density is
 \be 
 n_{0}(x) \equiv \langle \psi_0(N) | \hat\phi^{\dag}(x)\hat\phi(x)| \psi_0(N)\rangle= 2 n \sin^2 (\pi x /L). 
\ee 
The most visible effect due to the hard-wall trap  is to constrain 
the bosonic cloud in such a way that its density distribution presents a strong inhomogeneity which is the main physical 
difference compared to the periodic setup of Ref. \cite{kcc14}.

 \subsection{The quench protocol}
 
At time $t=0$ we suddenly turn on an infinitely strong interaction, i.e. we let the system evolve 
with the Hamiltonian (\ref{HLL}) with $c=\infty$. 
In this limit, known as Tonks-Girardeau limit \cite{TG},  the bosons behave as impenetrable. 
The Hamiltonian can be rewritten in terms of  hard-core bosonic fields, $\hat\Phi(x)$, $\hat\Phi^{\dag}(x)$ which satisfy a hybrid algebra; 
they commute at different space points, otherwise they obey an effective Pauli principle 
(induced by the infinite repulsion) whenever they are evaluated at the same space point: 
\begin{equation}
[\hat{\Phi}(x),\hat{\Phi}^\dag(y)]=0,\,x\neq y, \qquad[\hat{\Phi}^{\dag}(x)]^2=[\hat{\Phi}(x)]^2=0.
\label{alg2}
\end{equation} 
In terms of these fields the Hamiltonian is quadratic
\be\label{H_HCB}
\hat H = \int_{0}^{L} dx \, \partial_x \hat\Phi^{\dagger} (x) \partial_x \hat\Phi (x),
\ee
and the hybrid commutation relations encode the infinitely strong interactions which seem absent 
from the quadratic form (\ref{H_HCB}). 
The relation between the hard-core boson fields and the free bosonic ones is 
$\hat{\Phi}^{\dag}(x)=P_x\hat{\phi}^{\dag}(x)P_x$, 
where $P_x=|0\rangle\langle0|_x+|1\rangle\langle1|_x$ is the local projector on the truncated Hilbert space
with at most one boson at the point $x$. 
 
Using a Jordan-Wigner transformation, we can map the hard-core boson fields to fermion fields
\be\label{JW}
\hat{\Psi}(x)=\textnormal{exp}\left\{i\pi\int_0^xdz\hat{\Phi}^{\dag}(z)\hat{\Phi}(z)\right\}\hat{\Phi}(x),\qquad\hat{\Psi}^{\dag}(x)=\hat{\Phi}^{\dag}(x)\textnormal{exp}\left\{-i\pi\int_0^xdz\hat{\Phi}^{\dag}(z)\hat{\Phi}(z)\right\},
\ee
which satisfy canonical anti-commutation relations $\{\hat\Psi(x),\hat\Psi^{\dag}(y)\}=\delta(x-y)$. 
The Jordan-Wigner mapping guarantees that the fermionic and the bosonic density operators coincide, i.e 
$\hat\Psi^{\dag}(x)\hat\Psi(x) = \hat\Phi^{\dag}(x)\hat\Phi(x)$.

In terms of the fermionic fields the Hamiltonian (\ref{H_HCB}) is 
\be\label{H_Fermi}
\hat H = \int_{0}^{L} dx \, \partial_x \hat\Psi^{\dagger} (x) \partial_x \hat\Psi (x),
\ee
which is diagonalised by the Fermi operators $\hat\eta_{q}$, $\hat\eta^{\dag}_{q}$, related to the fermionic fields $\hat\Psi(x)$, 
$\hat\Psi^{\dag}(x)$ as
\be\label{psi_eta}
\hat{\Psi}(x)=\sum_{q=1}^\infty \varphi_q(x) \hat\eta_q, \quad \hat\eta_q=\int_0^L dx \, \varphi^{*}_q(x)\hat{\Psi}(x),
\ee
where the post-quench single-particle eigenfunctions coincide with the pre-quench single-particle ones in Eq. (\ref{eigenfunctions}). 
The crucial difference between the two set of modes is  the different algebra they satisfy.
In terms of the fermionic modes the Hamiltonian is diagonal
\begin{equation}
\hat H = \sum_{q=1}^\infty \epsilon_q \,\hat\eta^{\dag}_q \hat\eta_q=
\sum_{q=1}^\infty \epsilon_q \, \hat n_{q},
\end{equation}
with $\hat n_q \equiv \hat\eta^{\dag}_q \hat\eta_q$ being the post-quench mode occupation operators.

The main observable that we consider in the following is two-point fermionic correlation function
\be\label{fermi_corr_t}
C(x,y;t)\equiv %\langle \hat\Psi^{\dag}(x)\hat\Psi(y)\rangle_t =
\langle\exp(i \hat H t) \hat\Psi^{\dag}(x)\hat\Psi(y) \exp(-i \hat H t)\rangle,
\ee
where we introduced the simplified notation $\langle\dots\rangle \equiv \langle\Psi_0(N)|\dots|\Psi_0(N)\rangle$ 
in order to indicate expectation values in the initial state.
The time dependence in this correlation function can be explicitly written in terms of the post-quench modes as
\be\label{fermi_corr_t_eta}
C(x,y;t) = \sum_{p,q}\varphi^*_p(x)\varphi_q(y){\rm e}^{i(\epsilon_p-\epsilon_q)t}
\langle\hat\eta^{\dag}_p \hat\eta_q \rangle.
\ee
The density profile is just given by the correlation function evaluated at coincident points
\be
n(x;t)= C(x,x;t)=\sum_{p,q}\varphi^*_p(x)\varphi_q(x){\rm e}^{i(\epsilon_p-\epsilon_q)t}
\langle\hat\eta^{\dag}_p \hat\eta_q \rangle.
\label{nxtdef}
\ee

\section{Stationary properties}
\label{sec3}

In this section we report a complete characterisation of the stationary properties of the system after the quench. 
We compute the infinite time average of the density profile and of fermionic correlation function 
which equal their large time limit, as we will explicitly show only in the following section.  

Because of the integrability of the post-quench Hamiltonian, the large time limit of the reduced density matrix of 
any finite interval (in the sense described in Refs. \cite{cdeo-08,bs-08,CEF,CEFII}) is expected to be described by the GGE \cite{gg}
\be
\rho_{GGE}= Z^{-1} \exp\left(-\sum_i \lambda_i \hat I_i\right), 
\ee
where $\{\hat I_i\}$ is a complete set of local integrals of motion and the Lagrange multipliers $\lambda_i$ are fixed by the 
conditions $\langle \hat I_i \rangle={\rm Tr} [\rho_{GGE}\hat I_i]$. 
However, recent results \cite{f-14,noGGE1,noGGE2,noGGEp,noGGE3,noGGE4}
show that in some interacting theories the stationary state differs from the GGE built with local charges \cite{fe-13b,fcce-13},
suggesting that additional integrals of motion should be included in the GGE.
Here we can ignore this issue since we are dealing with a post-quench free theory.
Furthermore, we also prefer to avoid dealing with the issue of locality because our post-quench Hamiltonian breaks translational invariance. 
Thus we exploit the fact that the Tonks-Girardeau model has a simpler infinite set of conserved charges, 
formed by the fermionic mode occupation numbers $\hat n_q$ (we recall that for PBC the local conserved charges can be expressed as 
linear combinations of $\hat n_q$ \cite{fe-13,csc13}, so the GGE's built from $\hat n_q$ and $\hat I_i$ are equivalent). 

The time average of Eq. (\ref{fermi_corr_t_eta}) can be straightforwardly worked out 
(see also \cite{sc-14} for more general settings) obtaining 
\be\label{C_GGE_def}
 \overline{C(x,y;t)}  = \sum_{q=1}^\infty \varphi^*_q(x)\varphi_q(y) \langle\hat n_q \rangle =\lim_{t\to\infty}C(x,y;t) \equiv C_{\infty}(x,y) ,
\ee
which, as expected, only depends on the post-quench fermionic mode occupation.
We emphasise that $\langle\hat n_q \rangle$ is the only needed ingredient to construct the GGE 
and, thanks to Wick's theorem, it allows us to calculate any correlation function of local operators, 
showing that the GGE indeed captures the complete stationary behaviour.
It is worth mentioning that the GGE also fixes stationary two-time quantities \cite{eef-12}, 
which however will not be considered here.

Thus the elementary bricks needed  for the stationary (\ref{C_GGE_def}) and time-dependent (\ref{fermi_corr_t_eta})
fermionic correlation function are the fermionic mode occupation 
$\langle \hat n_q\rangle$ and the correlator $\langle \hat \eta^{\dag}_p \hat\eta_q \rangle$, respectively. 
In analogy to the periodic case \cite{kcc14}, these can be obtained from the initial correlator of the real-space fermionic fields 
$\langle\hat\Psi^{\dag}(x)\hat\Psi(y)\rangle$ calculated in the next subsection.

\subsection{Initial fermionic correlation function}

\begin{figure}[t]
\includegraphics[width=0.5\textwidth]{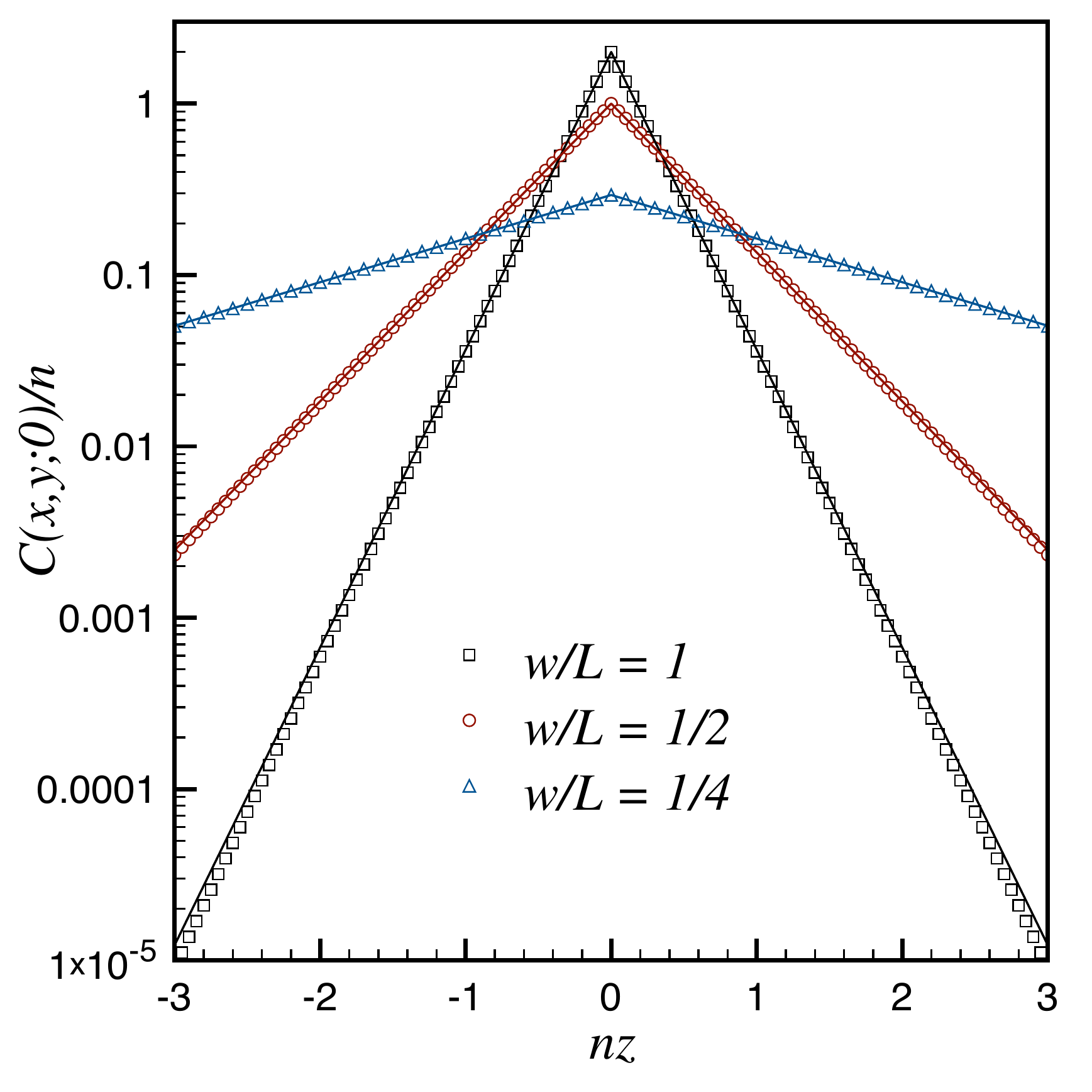}
\caption{The initial fermionic correlation function Eq. (\ref{PsiPsi_0_v2}) as function of $z=x-y$ for fixed $w=x+y$.
The numerical data for $N=L= 200$ (symbols) are compared to the analytical scaling function given by Eq. (\ref{PsiPsi_0_TDL}) (full lines). 
}
\label{C00}
\end{figure}

The calculation of the initial fermionic correlation function is not straightforward because of  
the non-Gaussian nature of the initial state in terms of the post-quench fermionic operators. 
As a starting point, we should then exploit the usual relation between fermionic and bosonic correlation which reads (for $x<y$)
 \be\label{PsiPsi_0}
 \langle\hat\Psi^{\dag}(x)\hat\Psi(y)\rangle=
 \sum_{j=0}^{\infty}\frac{(-2)^j}{j!}\int_{x}^{y}dz_1\ldots\int_{x}^{y}dz_j
 \langle\hat\Phi^{\dag}(x)\hat\Phi^{\dag}(z_1)\ldots\hat\Phi^{\dag}(z_j)\hat\Phi(z_j)\ldots \hat\Phi(z_1)\hat\Phi(y)\rangle.
 \ee
Although the initial state does not respect the hard-core condition, we can treat the hard-core boson fields as if they were canonical 
bosonic fields. Indeed following the analogous idea for PBC \cite{kcc14}, we can assume
 \be\label{HCtoBoson}
 \langle \hat\phi^{\dag}(x)\hat\phi^{\dag}(z_1)\ldots\hat\phi^{\dag}(z_j)\hat\phi(z_j)\ldots \hat\phi(z_1)\hat\phi(y)\rangle
= \langle \hat\Phi^{\dag}(x)\hat\Phi^{\dag}(z_1)\ldots\hat\Phi^{\dag}(z_j)\hat\Phi(z_j)\ldots\hat\Phi(z_1)\hat\Phi(y)\rangle.
 \ee
This equality is proved in the Appendix \ref{appA} using a rigorous lattice regularisation.
The lhs of Eq. (\ref{HCtoBoson}) is straightforwardly worked out in the initial ground state  $|\Psi_0(N)\rangle$:
 \be
 \langle \hat\phi^{\dag}(x)\hat\phi^{\dag}(z_1)\ldots\hat\phi^{\dag}(z_j)\hat\phi(z_j)\ldots \hat\phi(z_1)\hat\phi(y)\rangle
 =\varphi^{*}_1(x)\varphi_1(y) \, \prod_{i=1}^{j}|\varphi_1(z_i)|^2
 \langle (\hat \xi^{\dag}_1)^{ j+1} (\hat \xi_1)^{j+1}\rangle.
 \label{string}
 \ee
From $\hat\xi_{1} |\Psi_0(N)\rangle=\sqrt{N}|\Psi_0(N-1)\rangle$,
we have $\langle (\hat \xi^{\dag}_1)^{ j+1} (\hat \xi_1)^{j+1}\rangle = N!/(N-j-1)!$, which allows us to rewrite
Eq. (\ref{PsiPsi_0})  as
\be
 \langle\hat\Psi^{\dag}(x)\hat\Psi(y)\rangle
 =\varphi^{*}_1(x)\varphi_1(y)\sum_{j=0}^{\infty}\frac{(-2)^j}{j!}\frac{N!}{(N-j-1)!}\left(\int_x^y dz |\varphi_1(z)|^2\right)^j.
\ee
This relation is valid in the domain $x<y$, while in the opposite case $x>y$, the only difference arises from the 
exchange of the integration limits, leading to the absolute value of the integral which can be written as  
\bea
\left | \int_x^y dz |\varphi_1(z)|^2  \right |
& = &\frac{|x-y|}{L}+\frac{{\rm sgn}(x-y)}{2\pi}\left[\sin\left(\frac{2\pi y}{L}\right)-\sin\left(\frac{2\pi x}{L}\right)\right] \nonumber \\
& = & \frac{|x-y|}{L}-\frac{{\rm sgn}(x-y)}{\pi}\cos\left[\frac{\pi (x+y)}{L}\right]\sin\left[\frac{\pi(x-y)}{L}\right],
\eea
which finally leads to ($\forall \, x,y\in[0,L]$)
\be\label{PsiPsi_0_v2}
 \langle\hat\Psi^{\dag}(x)\hat\Psi(y)\rangle 
 = 2 \frac{N}{L} \sin\left(\frac{\pi}{L}x\right)\sin\left(\frac{\pi}{L}y\right)
 \left[1-2\left(\frac{|x-y|}{L}-\frac{{\rm sgn}(x-y)}{\pi}\cos\left(\frac{\pi (x+y)}{L}\right)\sin\left(\frac{\pi(x-y)}{L}\right)\right)\right]^{N-1}.
\ee
Eq. (\ref{PsiPsi_0_v2}) is valid for any finite value of $L$ and $N$.
Interestingly, its structure is quite general and independent from the particular shape of the 
confining potential (see Appendix \ref{appB} for more details).

As we shall see, a relevant scaling regime in the quench problem is provided by taking the thermodynamic limit (TDL) with 
$\tilde w=(x+y)/L$ kept fixed and $z=x-y$ arbitrary. In this limit, Eq. (\ref{PsiPsi_0_v2}) becomes
\be\label{PsiPsi_0_TDL}
 \langle\hat\Psi^{\dag}(x)\hat\Psi(y)\rangle 
 = n\,[1-\cos(\pi\tilde w)] e^{ -2 n [1-\cos(\pi \tilde w)] |z| }.
 \ee
In Fig. \ref{C00}, this asymptotic form is compared with the direct numerical evaluation of Eq. (\ref{PsiPsi_0_v2}).
In the following sections we will always use as a starting point Eq.  (\ref{PsiPsi_0_v2}), even if in some instances 
Eq. (\ref{PsiPsi_0_TDL}) would have led to the same results. 
We preferred to proceed in this way to keep the information about the boundaries as much as possible.

\subsection{Fermionic mode occupation}

 %%%%%%% FIGURE n_q %%%%%%%%
 \begin{figure}[t]
\includegraphics[width=0.5\textwidth]{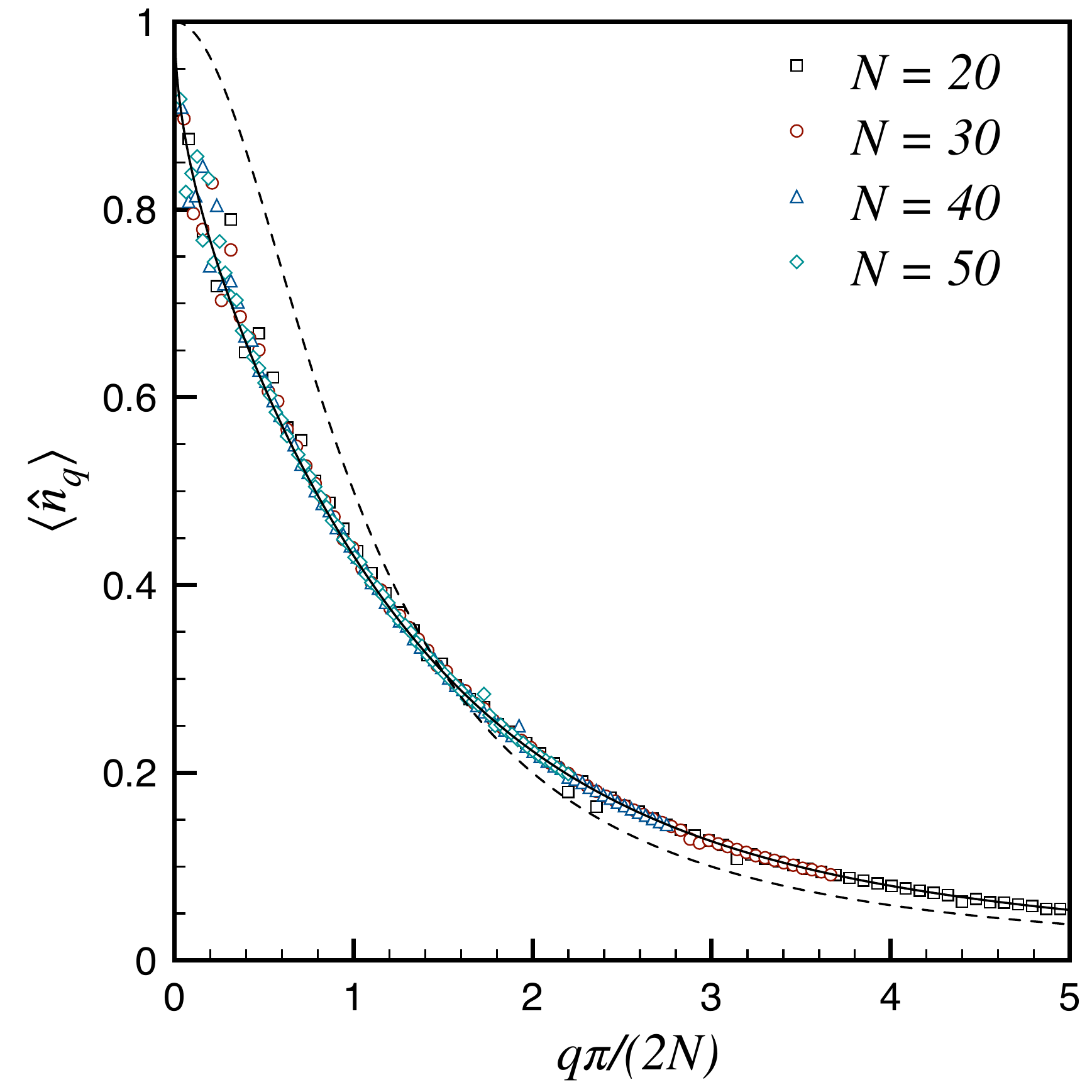}
\caption{Fermionic mode occupation $\langle \hat n_{q} \rangle$ as function of 
the rescaled variable $q\pi/(2N)$ for different particle numbers $N$. 
The numerical data, evaluated using Eq. (\ref{nq_int1}), collapse on the asymptotic universal function (full black line)
given by Eq. (\ref{nq_TDL}). 
For comparison we show (dashed line) the mode occupation for PBC \cite{kcc14}.} 
\label{fig1}
\end{figure}
 %%%%%%%%%%%%%%%%%%%%%% 

The fermionic mode occupation $\langle\hat n_{q}\rangle$ is obtained plugging Eq. (\ref{PsiPsi_0_v2}) in the definition  (\ref{psi_eta}):
\bea
 \langle \hat n_q \rangle 
 & = &\int_0^L dx \int_0^Ldy \, \varphi_q(x)\varphi^{*}_q(y) \langle \hat\Psi^{\dag}(x)\hat\Psi(y)\rangle \nonumber \\
 & = & \frac{4N}{L^2} \int_0^Ldx\int_0^Ldy \, \sin\left( \frac{q\pi x}{L} \right) \sin\left( \frac{q\pi y}{L} \right)
  \sin\left( \frac{\pi x}{L} \right)  \sin\left( \frac{\pi y}{L} \right) \nonumber \\
 & & \times \,   \left[1-2\left(\frac{|x-y|}{L}-\frac{{\rm sgn}(x-y)}{\pi}\cos\left(\frac{\pi (x+y)}{L}\right)\sin\left(\frac{\pi(x-y)}{L}\right)\right)\right]^{N-1}.
\eea
Using standard trigonometric identities and changing
the integration variables to $v = \pi (x-y) /L $ and $u = \pi (x+y)/L-\pi$, we have the more compact expression
\be\label{nq_int1}
  \langle \hat n_q \rangle = \frac{N}{2\pi^2}\int_{-\pi}^{\pi}du\int_{|u|-\pi}^{\pi-|u|}dv \,
  [ \cos(qv)-(-1)^{q}\cos(qu) ] [\cos(v)+\cos(u)]
  \left[1-\frac{2}{\pi} [|v|+{\rm sgn}(v)\sin(v)\cos(u)]\right]^{N-1}. 
 \ee
In Fig. \ref{fig1} we report the mode occupation $\langle \hat n_{q} \rangle$  evaluated numerically from Eq. (\ref{nq_int1}) 
for different values of $N$. 
All data for different $N$ collapse on a universal smooth function of the rescaled variable $q/N$,
except for very small values of $q$ ($q \lesssim 10$).
This implies that only modes with $q/N\sim O(1)$ are important in the TDL allowing us to 
simplify Eq. (\ref{nq_int1}).
Indeed, for $q\gg 1$, the function $\cos(q u)$ is integrated over an integer multiple of its period (we recall that $q$ is an integer) 
and consequently its contribution is suppressed with respect to the remaining part of the integral. 
Thus, for large $q$ and large $N$, we can rewrite Eq. (\ref{nq_int1}) as
\be \label{nqstart}
  \langle \hat n_q \rangle = \frac{2N}{\pi^2}\int_{0}^{\pi}du\int_{0}^{\pi-u}dv \,
  \cos(qv) [\cos(v)+\cos(u)]  \left[1-\frac{2}{\pi} [v+\sin(v)\cos(u)]\right]^{N-1}. 
 \ee
We can now take the large $N$ limit. Since  $ \left|1-\frac{2}{\pi} [v+\sin(v)\cos(u)]\right| \le 1$
throughout the integration domain, for large $N$, the integral in $v$ is dominated by the neighbourhood of $v=0$
and so we can limit the integral to a region $v\in[0,\epsilon]$ with $\epsilon \ll 1$. 
Expanding in $v$ the integrated function, we obtain
\be
  \langle \hat n_q \rangle = \frac{2N}{\pi^2}\int_{0}^{\pi}du\int_{0}^{\epsilon}dv \, 
  \cos(qv) [1+\cos(u)] 
  \left[1-\frac{2v}{\pi} [1+\cos(u)]\right]^{N-1}, 
\ee
where we could not expand $\cos(q v)$ because for large $q$ it can oscillate many times in $[0,\epsilon]$. 
For large $N$ and small $v$ it holds
\be  
\Big[1-\frac{2v}{\pi} [1+\cos(u)]\Big]^{N-1} \simeq e^{ N\ln\big(1-\frac{2v}{\pi} [1+\cos(u)] \big) }
\simeq e^{-  N\frac{2v}{\pi} [1+\cos(u)] },
\ee
leading to
\be
  \langle \hat n_q \rangle = \frac{2N}{\pi^2}\int_{0}^{\pi}du\int_{0}^{\epsilon}dv \,
  \cos(qv) [1+\cos(u)] 
  {\rm e}^{-N\frac{2v}{\pi} [1+\cos(u)]}.
\ee 
After these simplifications, the integrated function is exponentially small in $N$, and therefore, in
the TDL, we can send the upper bound of integration $\epsilon$ to infinity.
The $v$ integration becomes  the cosine Fourier transform of the exponential function, finally giving
\bea\label{nq_TDL}
\langle \hat n_q \rangle 
& = & \frac{1}{\pi}\int_0^\pi   \frac{du}{1+\left[\frac{q\pi/(2N)}{1+\cos(u)}\right]^{2} }
 =  1-\sqrt{ \frac{\tilde q \, (\tilde q + \sqrt{4+\tilde q\,^{2}} )}{2\,(4+\tilde q\,^{2})} }, 
\qquad \tilde q \equiv \frac{q\pi}{2N} = \frac{1}{2n}\frac{q\pi}{L} ,
\eea
showing explicitly that the mode occupation number is indeed a function of the rescaled variable $q/N$.
This analytic result is compared in Fig. \ref{fig1} with the numerical evaluation of the mode occupation 
and they perfectly match for large enough $N$. Notice that Eq. (\ref{nq_TDL}) satisfies the normalisation condition
$\sum_{q} \langle \hat n_q \rangle = N$ in the TDL, as it should.

Before using this result to calculate real-space properties of the system in the stationary state,   
it is interesting to compare this mode distribution with the same quantity for PBC 
$n_{\rm PBC}(\tilde q) = 1/(1+\tilde q\,^{2})$ \cite{kcc14} (reported for comparison as a dashed line in Fig. \ref{fig1}). 
Both distributions have a power law behavior for $\tilde q \gg 1$ with  $n_{\rm PBC}(\tilde q) \simeq 1/\tilde q\, ^{2}$,
and $\langle \hat n_q \rangle \simeq 3/(2 \tilde q\,^{2})$.
The effect of the boundaries is more apparent for small $\tilde q$ when 
$n_{\rm PBC}(\tilde q) \simeq 1- \tilde q\, ^{2}$ and  $\langle \hat n_q \rangle \simeq 1- \sqrt{\tilde q}/2$, 
the latter being non-analytic in zero (which, as we shall see, has strong consequences for real space correlations).

%%%%%%% FIGURE n_GGE %%%%%%%%
 \begin{figure}[t]
\includegraphics[width=0.5\textwidth]{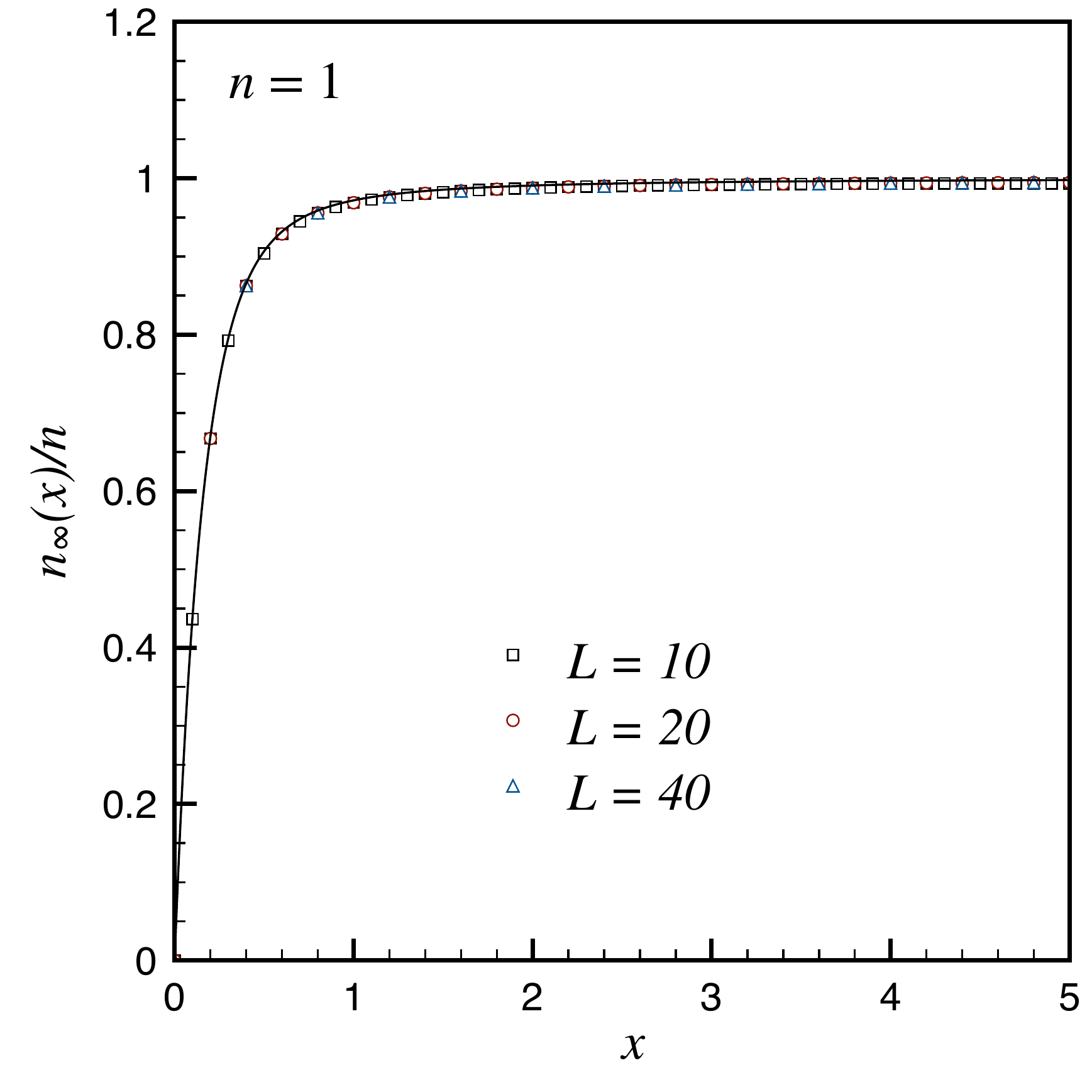}\includegraphics[width=0.5\textwidth]{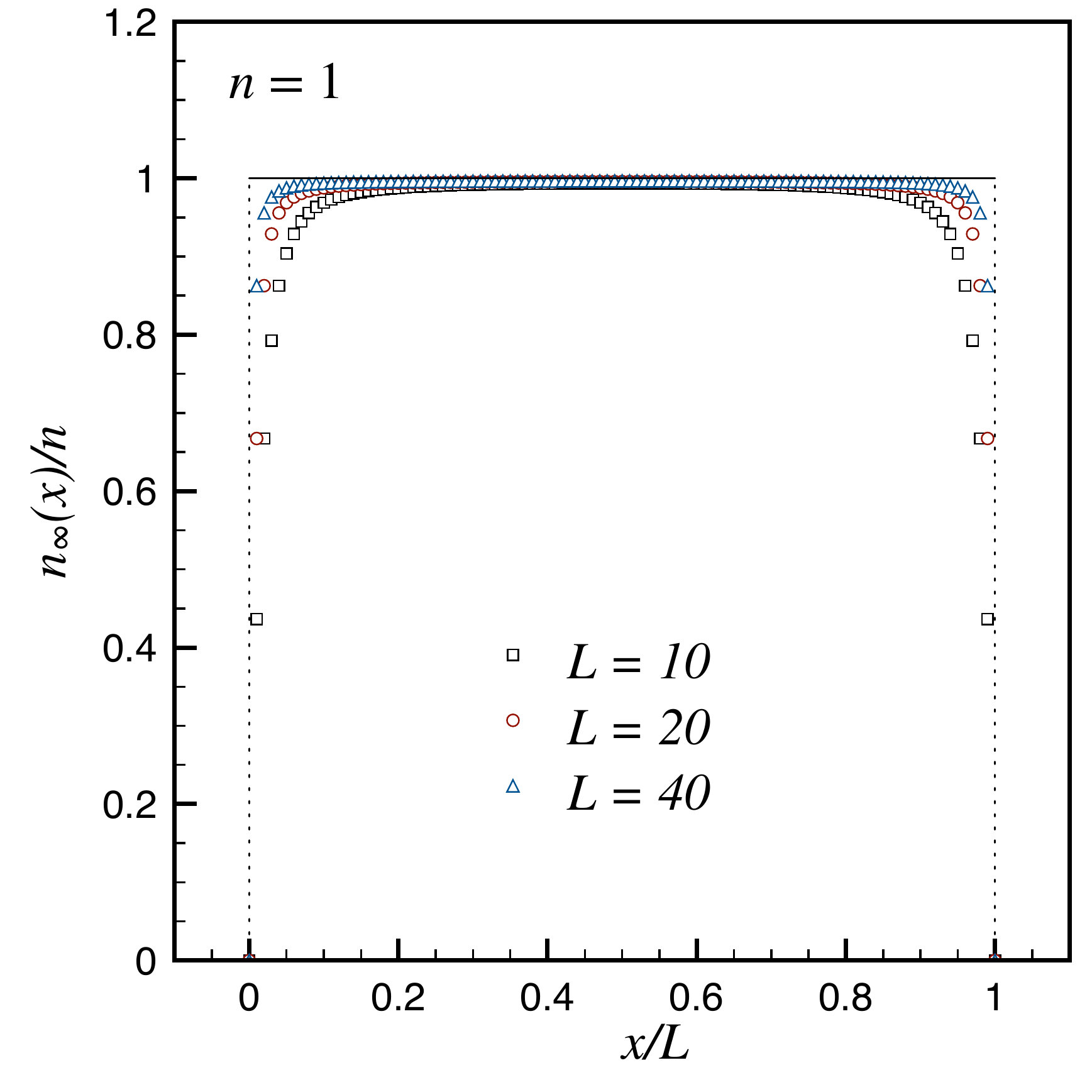}
\caption{(Left) The stationary particle density profile near the left boundary ($x=0$) is perfectly described by Eq. (\ref{nGGE}).
(Right) The same data vs the rescaled variable $x/L$ showing that for large $L$, the systems becomes homogeneous 
in the interval $[0,L]$ with small corrections at the boundaries. } 
\label{fig2}
\end{figure}
 %%%%%%%%%%%%%%%%%%%%%%

\subsection{Stationary particle density}
In this section we analyse the particle density profile $n_{\infty}(x)$ in the stationary state. 
From the definition (\ref{C_GGE_def}), using the one-particle eigenfunctions (\ref{eigenfunctions})
and the fermionic mode occupation (\ref{nq_TDL}), one immediately has
\bea
n_{\infty}(x) 
& = & \sum_{q=1}^\infty |\varphi_q(x)|^{2} \langle \hat n_{q} \rangle 
 =    \frac{2}{\pi L} \sum_{q=1}^\infty \int_0^\pi du 
\frac{\sin^2(q\pi x /L)}{1+\left[\frac{q\pi/(2N)}{1+\cos(u)}\right]^{2}}.
\eea
In Fig. \ref{fig2}, we report the numerical evaluation of this sum showing that it approaches the uniform value $n=N/L$
when increasing the system size $L$ and particle number $N$. 
The HBC influence the profile only in a region $x\sim O(1)$ close to the boundaries 
which shrinks to a set of zero measure when considering the scaling variable $x/L$ (right panel). 

This behaviour can be easily understood analytically by replacing, in the TDL limit, the sum with an integral
\be
n_{\infty}(x) =  \frac{2}{\pi^2} \int_{0}^{\infty} dq \int_{0}^{\pi} du 
\frac{\sin^2(xq)}{1+\left[\frac{q/(2n)}{1+\cos(u)}\right]^{2}}=
n_{\infty}(x) = n - n {\rm e}^{-4nx}[{\rm I}_{0}(4nx)-{\rm I_1}(4nx)],
\label{nGGE}
\ee
where ${\rm I}_{m}(z)$ are the modified Bessel functions. 
This shows that the thermodynamic stationary density is $n_\infty(x)= n$, i.e. the value obtained 
in the rescaled variable $x/L$ (see the right panel in Fig. \ref{fig2}).
The correction in Eq. (\ref{nGGE}) is non vanishing only in a set of measure zero (in $x/L$)
and  describes the behaviour close to the left boundary at $x=0$ (which perfectly matches 
the numerical result as shown in the left panel of Fig. \ref{fig2}).
Notice that in Eq. (\ref{nGGE}) we lost the information about the right boundary at $x=L$. 
It is however obvious that close to the right boundary the density has the same profile as at the left one.

%%%%%%% FIGURE C_GGE %%%%%%%%
 \begin{figure}[t]
\includegraphics[width=0.5\textwidth]{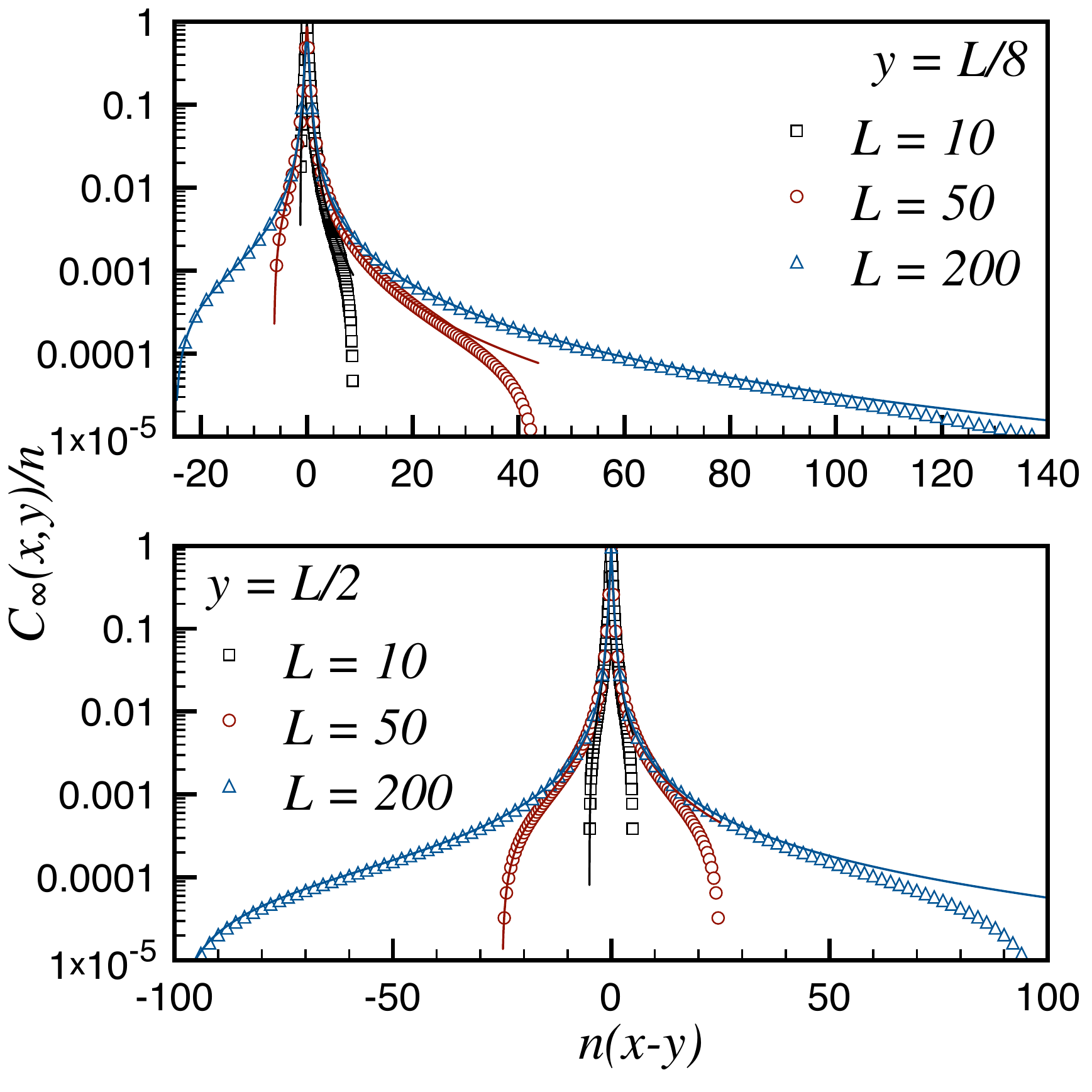}\includegraphics[width=0.5\textwidth]{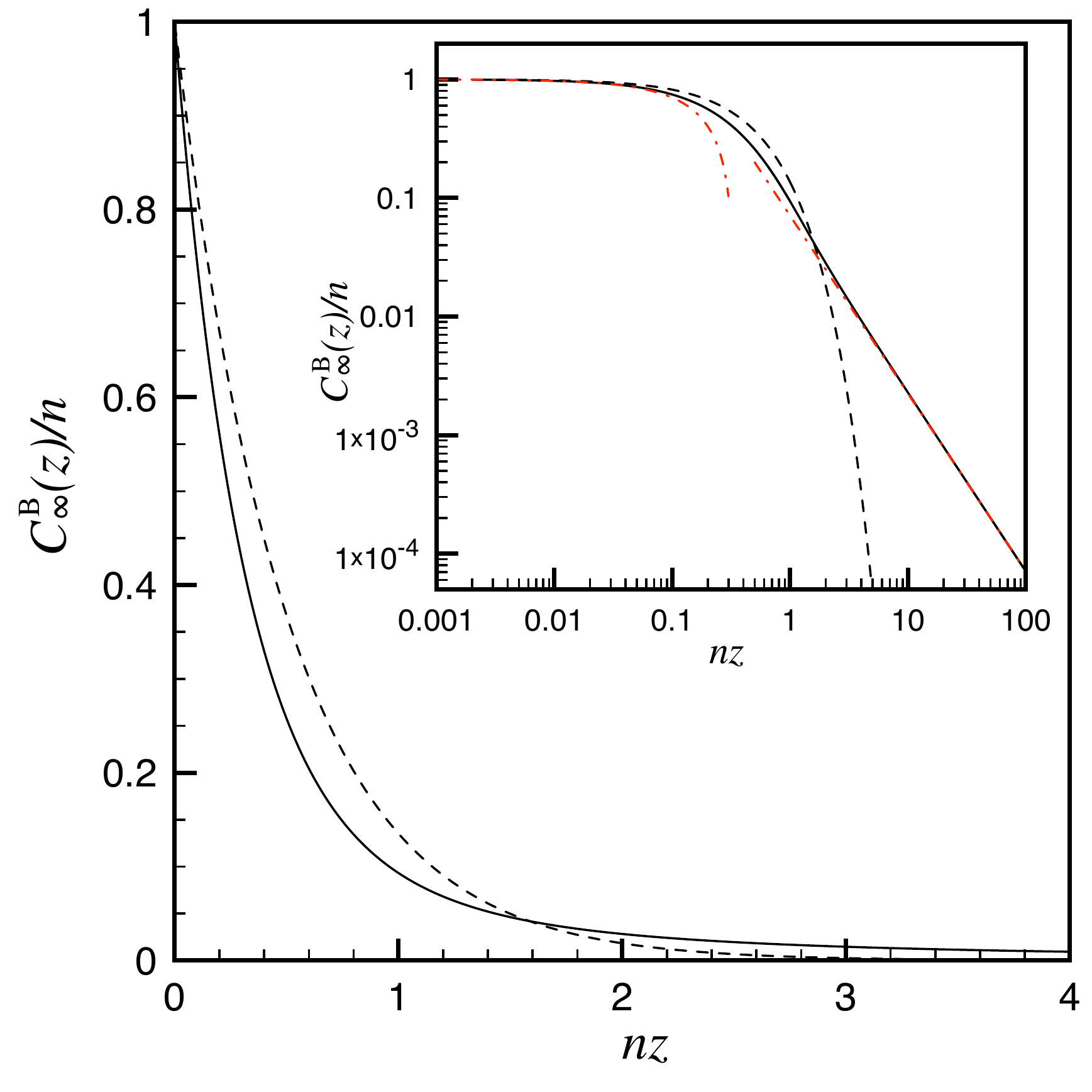}
\caption{Fermionic two-point correlation function in the stationary state.
(Left) The numerical evaluated correlators (using the sum in Eq. (\ref{CGGE_N}))  
are compared to the TDL result in Eq. (\ref{CGGE_TDL}) (full lines)
for $y=L/2$ (bottom) and $y=L/8$ (top).
(Right) The bulk stationary correlator in Eq. (\ref{CGGE_TI}) (full black lines) is compared to the 
PBC one (dashed black lines).
In the inset, the same correlators are shown in log-log scale to highlight the power law behaviour. 
The red dot-dashed lines represent the asymptotic behavior  for small $z$ (i.e. $\sim 1-3n|z|$) 
and for large $z$ (i.e. $\sim |nz|^{-3/2}/(8\sqrt{\pi})$). 
} 
\label{fig3}
\end{figure}
 %%%%%%%%%%%%%%%%%%%%%%

\subsection{Stationary two-point fermionic correlation function}

In this section we study the two-point fermionic correlator $C_{\infty}(x,y)$ in the stationary state.
Let us start by noticing that as long as we are interested in bulk properties of the system,   
since the stationary density is homogeneous, we expect all correlation functions to be translational invariant. 
Therefore, we keep the difference $x-y \sim O(1)$  to avoid infinitely separated points in the TDL.

In terms of the mode occupation, the stationary fermionic correlation function can be written as
\bea\label{CGGE_N}
C_{\infty}(x,y) 
& = & \sum_{q=1}^\infty \varphi^{*}_q(x) \varphi_q(y) \langle \hat n_{q} \rangle 
 =    \frac{2}{\pi L} \sum_{q} \int_0^\pi du 
\frac{\sin(q\pi x /L)\sin(q\pi y /L)}{1+\left[\frac{q\pi/(2N)}{1+\cos(u)}\right]^{2}}\\
& = &  \frac{1}{\pi L} \sum_{q=1}^\infty \int_0^\pi du
\frac{\cos[q\pi (x-y) /L]-\cos[q\pi (x+y) /L]}{1+\left[\frac{q\pi/(2N)}{1+\cos(u)}\right]^{2}},\nonumber
\eea
that, in the TDL, becomes a double integral which can be explicitly performed 
\bea\label{CGGE_TDL}
C_{\infty}(x,y) 
& = & \frac{1}{\pi^2} \int_{0}^{\infty} dq \int_{0}^{\pi} du
 \frac{\cos[q (x-y)]-\cos[q (x+y) ]}{1+\left[\frac{q/(2n)}{1+\cos(u)}\right]^{2}} \nonumber\\
 & = & n {\rm e}^{-2n |x-y|}[{\rm I}_{0}(2n|x-y|)-{\rm I_1}(2n|x-y|)]
 - n {\rm e}^{-2n |x+y|}[{\rm I}_{0}(2n|x+y|)-{\rm I_1}(2n|x+y|)].
\eea
This stationary correlator consists of two different parts: 
\be
C_{\infty}(x,y) = C^B_{\infty}(|x-y|) +C_{\infty}^{\rm bou}(x,y) ,
\ee
where we have (i) a bulk correlator $C^B_{\infty}(|x-y|)$ depending only on the distance between the two points and 
which is the true thermodynamic stationary correlator, and 
(ii) a boundary term $C_{\infty}^{\rm bou}(x,y)$ depending on $x+y$ which goes to zero when 
$x$ and $y$ are far from the boundary $x,y\gg1$. 
We stress that the information about the right boundary has been lost because of the way we performed the TDL, 
in analogy to the density profile.

In the two left panels of Fig. \ref{fig3} we compare the numerically evaluated correlation function 
with the thermodynamic result for $y=L/2,\,L/8$ as a function of $x\in[0,L]$. 
The numerics perfectly agree with  Eq. (\ref{CGGE_TDL}) as far as
$x$ is far from the right boundary where, as we already stated, Eq. (\ref{CGGE_TDL}) does not apply.

Let us discuss in more details the bulk stationary correlator 
\be\label{CGGE_TI}
C_{\infty}^B(z) 
= n {\rm e}^{-2n |z|}[{\rm I}_{0}(2n|z|)-{\rm I_1}(2n|z|)]. 
\ee
Although this result slightly resemble the real space correlator for PBC $C_{\rm PBC}(z)=n {\rm e}^{-2n |z|}$ \cite{kcc14},
they are qualitatively different as evident from Fig. \ref{fig3}.
The boundary conditions and the highly inhomogeneous initial profile strongly affect the two-point stationary function {\it in the  bulk}. 
The multiplicative factor in Eq. (\ref{CGGE_TI}), depending on the difference between two Bessel functions, 
modifies both the small- and the large-distance behaviour. 
For $zn\ll 1$ the correlator behaves as $C_{\infty}^B(z)/n \sim 1-3n|z|$ manifesting a faster short-distance decay compared to the  
PBC case. 
For large distances the behaviour is completely different.  
While for PBC there is exponential decay for all distances, the trapped correlator shows an algebraic decay for $zn \gg 1$, 
namely $C_{\infty}^B(z)/n \sim |n z|^{-3/2}/(8\sqrt{\pi})$
\footnote{From the mathematical point of view the algebraic decay in $z$ is simply a consequence of the fact 
that its Fourier transform $\langle \hat n_{q}\rangle$ is not analytic in zero}.
This is actually a very important  difference compared to the PBC case: in the TDL, 
the boundaries strongly affect the bulk, a phenomenon that has no direct analogue in finite temperature systems.

\subsection{Stationary bosonic correlation function}

%%%%%%% FIGURE C_B %%%%%%%%
 \begin{figure}[t]
\includegraphics[width=0.5\textwidth]{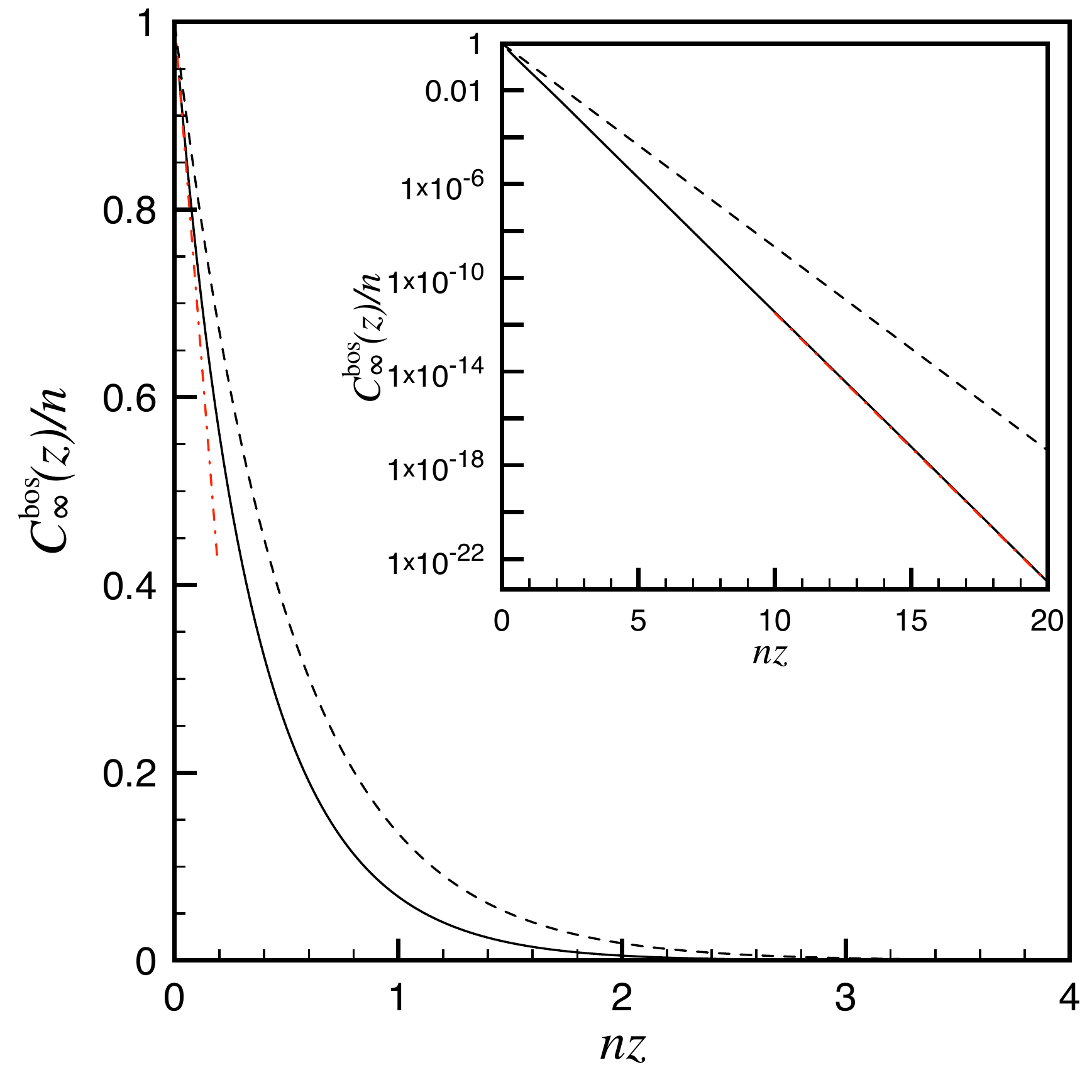}
\caption{Bosonic bulk two-point correlation function in the stationary state. 
For small distances $nz\ll 1$, $C^{\rm bos}_\infty(z)/n\simeq 1-3 n |z|$ (dot-dashed red line). 
The inset shows the same correlator in log-scale to highlight the exponential large distance behaviour 
$C^{\rm bos}_\infty(z)/n\propto e^{-2.65 n |z|}$ (dot-dashed red line). 
In both panels, the dashed lines correspond to the bosonic correlator for PBC which is shown for comparison. 
} 
\label{fig3CB}
\end{figure}
 %%%%%%%%%%%%%%%%%%%%%%

From the knowledge of the fermionic two-point function all other correlations in the stationary state can be derived 
with the help of Wick's theorem. 
The most relevant correlation from the experimental point of view is surely the bosonic two-point correlation function
whose Fourier transform is the momentum distribution function, the quantity most commonly measured in cold atoms experiments.
In the remaining of this section we are going to evaluate this bosonic correlation function in the bulk. 

We consider the stationary bosonic two-point correlation function
\be
C^{\rm bos}_\infty(z)\equiv \lim_{t\to\infty} \langle\hat{\Phi}^\dagger(x,t)\hat{\Phi}(x+z,t) \rangle,
\ee
where we explicitly used bulk  translational invariance in the stationary state. In the following, we will set $x=0$  for simplicity.
The bosonic correlation function  
$C^{\rm bos}_\infty(z)$ can be expressed in terms of the fermionic correlations using 
the Jordan-Wigner mapping (\ref{JW}) and  Wick's theorem. Indeed, for $z>0$, we have
\begin{equation}
C^{\rm bos}_\infty(z) 
= \left\langle \hat{\Psi}^{\dag}(0)\,\exp\left\{-i\pi\int_{0}^{z}dy\,\hat{\Psi}^{\dag}(y) \hat{\Psi}(y)\right\}  \hat{\Psi}(z) \right\rangle.
\end{equation}
Taylor expanding the exponential this becomes
\be
C^{\rm bos}_\infty(z)   =  \sum_{k=0}^{\infty}\frac{(-i\pi)^k}{k!}\int_{0}^{z}dz_{1}\cdots\int_{0}^{z}dz_{k}
\langle \hat{\Psi}^{\dag}(0) \hat{\Psi}^{\dag}(z_{1})\hat{\Psi}(z_1)\cdots  \hat{\Psi}^{\dag}(z_{k})\hat{\Psi}(z_k)  \hat{\Psi}(z)\rangle,
\ee
which can be rearranged in normal order and, using Wick's theorem, we finally have 
\be\label{C_B_fredholm}
C^{\rm bos}_\infty(z)   =  
\sum_{k=0}^{\infty}\frac{(-2)^k}{k!}\int_{0}^{z}dz_{1}\cdots\int_{0}^{z}dz_{k}  \det_{ij} \langle \hat{\Psi}^{\dag}(x_{i})\hat{\Psi}(y_j) \rangle 
= \sum_{k=0}^{\infty}\frac{(-2)^k}{k!}\int_{0}^{z}dz_{1}\cdots\int_{0}^{z}dz_{k}  \det_{ij} C_\infty^{B}(x_i-y_j),
\ee
where the indices $i,j$ run from $0$ to $k$, and we used the convention 
$x_{i}=y_{i}\equiv z_{i},\,\forall i>0$, and $x_{0}\equiv 0,\,y_{0}\equiv z$. 
Eq. (\ref{C_B_fredholm})  is a Fredholm's minor of the first order \cite{fredholm}. 

It is in general very difficult to manipulate analytically Fredholm's minors and, 
for this reason, we decided to evaluate Eq. (\ref{C_B_fredholm}) numerically which is a quite standard procedure.
Indeed, this numerical evaluation can be achieved by discretising the Fredholm's minor in Eq. (\ref{C_B_fredholm}) 
as explained in Refs. \cite{adi,ksc-13,csc13}.
In order to do so, we proceed as follows: 
(i) we discretise the space interval $[0,z]$ in $M+1$ points, introducing the lattice spacing $a= z/(M+1)$; 
(ii) we define the $(M+1)\times (M+1)$ matrices (indices run form $1$ to $M+1$):
\bea 
\mathbb{R}_{nm}& = & \delta_{nm} - \delta_{n1}\delta_{1m},\\
\mathbb{S}_{nm}& = & C_\infty^{B}((n-m)a)\;{\rm for}\;n>1,\quad \mathbb{S}_{1m}  =   C_\infty^{B}(z-ma),\nonumber
\eea
where $C_\infty^{B}(z)$ is the stationary bulk fermionic correlation (\ref{CGGE_TI}).
Therefore, the bosonic correlator is given by the limit
\be
C^{\rm bos}_\infty(z) = \lim_{a\to 0} \frac{\det(2a\, \mathbb{S}-\mathbb{R})}{2a}.
\ee
In practice, we evaluate the ratio in the rhs of the above equation for small enough spacing $a$
and check that it does not vary to the required precision by making it smaller.
In this way, we numerically calculate $C^{\rm bos}_\infty(z)$ as a function of $z$  
and the results are reported in Fig. \ref{fig3CB}. 

Let us critically analyse the results in Fig. \ref{fig3CB}.
From the inset, it is clear that the large-distance behaviour of the bosonic correlation function is exponential, 
although the fermionic one is algebraic. 
This does not come as a surprise, because also in other cases \cite{csc13} the algebraic decay of fermionic correlations 
resulted in an exponential in the bosonic correlation. 
The decay rate of the exponential (i.e. the inverse correlation length) is $\sim 2.65 n $ which is larger than the decay rate 
for PBC $2 n$ (which is reported for comparison in Fig.  \ref{fig3CB}).
However, while for PBC the bosonic correlator is exactly exponential for all distances \cite{kcc14}, i.e. 
$C^{\rm bos}_{\rm PBC}(z)= n e^{-2 n|z|}$, this is not the case for HBC. 
Indeed for small $z$, $C^{\rm bos}_\infty(z)/n$ is well fitted by $1-3 n |z|$ (see Fig. \ref{fig3CB}), which coincides with
the small distance behaviour of the fermionic correlator.

\section{Time-dependent quantities}\label{sec4}

In this section we analyse the time evolution of the density profile and of the two-point fermionic correlation function.
We will limit to consider both these quantities deeply in the bulk in order to have more accessible results. 
In the periodic case these quantities are constant in time \cite{kcc14} because of translational invariance, but 
in the confined case they present a nontrivial dynamics. 
The inhomogeneous initial density affects for arbitrary times the non-equilibrium dynamics, leading 
for infinite time to the stationary bulk correlator given in Eq. (\ref{CGGE_TI}).
Indeed, we show that the time averaged values calculated in the previous section are indeed 
approached for long times in the TDL.

\subsection{The off-diagonal correlator $\langle \hat\eta^{\dag}_{p} \hat \eta_{q}\rangle$}
The elementary building block needed for the evaluation of the time-dependent quantities is the initial correlator of the post-quench
fermionic mode, which can be written for arbitrary $N$ and $L$ as 
\bea
 \langle \hat \eta^{\dag}_{p} \hat\eta_q \rangle 
 & = &\int_0^L dx \int_0^Ldy \, \varphi_p(x)\varphi^{*}_q(y) \langle \hat\Psi^{\dag}(x)\hat\Psi(y)\rangle \nonumber \\
 & = & \frac{4N}{L^2} \int_0^Ldx\int_0^Ldy \, \sin\left( \frac{p\pi x}{L} \right) \sin\left( \frac{q\pi y}{L} \right)
  \sin\left( \frac{\pi x}{L} \right)  \sin\left( \frac{\pi y}{L} \right) \nonumber \\
 & & \times \,   \left[1-2\left(\frac{|x-y|}{L}-\frac{{\rm sgn}(x-y)}{\pi}\cos\left(\frac{\pi (x+y)}{L}\right)\sin\left(\frac{\pi(x-y)}{L}\right)\right)\right]^{N-1}.
\eea
With the change of variables $v=\pi(x-y)/L$, $u=\pi(x+y)/L$, this can be rewritten as 
\be\label{etaeta_int1}
 \langle \hat \eta^{\dag}_{p} \hat\eta_q \rangle = \frac{N}{\pi^2}\int_{0}^{2\pi}du\int_{|u-\pi|-\pi}^{\pi-|u-\pi|}dv \,
  \sin[p(u+v)/2]\sin[q(u-v)/2]
   [\cos(v)-\cos(u)]
  \left[1-\frac{2}{\pi} [|v|-{\rm sgn}(v)\sin(v)\cos(u)]\right]^{N-1}. 
 \ee
We can proceed as in the case of the mode occupation, assuming that in the TDL the relevant contributions 
to the density and to the correlation functions only come from large $p$ and $q$ in Eq. (\ref{etaeta_int1}). 
Thus, we consider $p+q\gg 1$, but we make no assumption  about the difference $p-q$. 
%This assumption has been numerically checked and it turns out to be become exact in the TDL.

In order to make manifest the dependence on $p+q$ and $p-q$ of the above integral, let us use a few simple trigonometric identities. 
Let us start by expanding 
 \bea
 \sin[p(u+v)/2]\sin[q(u-v)/2] & = & [\cos(pv/2)\sin(pu/2) + \cos(pu/2)\sin(pv/2)]\nonumber \\
 & &\times  [ \cos(qv/2)\sin(qu/2) - \cos(qu/2)\sin(qv/2)].
 \eea
Then, let us focus the attention to one of the four products 
(the same argument will be valid for the other terms): 
\be
\cos(pv/2)\cos(qv/2)\sin(pu/2)\sin(qu/2) 
= \frac{1}{2} \cos(pv/2)\cos(qv/2) \{ \cos[u(p-q)/2] + \cos[u(p+q)/2] \}.
\ee
Now, for $p+q\gg 1$, since the integration domain in the variable $u$ always
contains an integer number of periods of the cosine function, we can neglect the 
term $\cos[u(p+q)/2]$. Collecting together the analogous results for all the four terms, one  gets
\be
\sin[p(u+v)/2]\sin[q(u-v)/2]  \simeq \frac{1}{2} 
\left\{ \sin[v(p+q)/2] \sin[u(p-q)/2] + \cos[v(p+q)/2]\cos[u(p-q)/2] \right\},
\ee
where the approximate equality is intended to be valid only under the integration in Eq. (\ref{etaeta_int1}) and in the TDL.

Therefore, making use of the fact that the integration domain in $v$ is symmetric  
(thus the term proportional to $\sin[v(p+q)/2] \sin[u(p-q)/2]$ vanishes identically) 
and changing the integration variable $u$ to $u-\pi$, we can straightforwardly recast Eq. (\ref{etaeta_int1}) into
\be\label{etaeta_int2}
\langle \hat \eta^{\dag}_{p} \hat\eta_q \rangle = \frac{N}{2\pi^2}\int_{-\pi}^{\pi}du\int_{|u|-\pi}^{\pi-|u|}\hspace{-2mm} dv 
\cos\big[v\frac{p+q}2\big]\cos\big[\frac{(u+\pi)(p-q)}2\big]
[\cos(v)+\cos(u)]
\Big[1-\frac{2}{\pi} [|v|+{\rm sgn}(v)\sin(v)\cos(u)]\Big]^{N-1}. 
\ee
Furthermore, whenever $p-q$ is odd, the argument of the integral is an odd function in the variable $u$ 
which integrated over the symmetric interval $[-\pi,\pi]$ gives zero. 

At this point, following the same reasoning leading from Eq. (\ref{nqstart}) to Eq. (\ref{nq_TDL}), we have 
\be\label{etaeta_TDL}
\langle \hat \eta^{\dag}_{p} \hat\eta_q \rangle =
\frac{1}{\pi}\int_0^\pi du  \, 
\frac{\cos[(u+\pi)(p-q)/2]}{1+\left[\frac{(p+q)\pi/(4N)}{1+\cos(u)}\right]^{2} } ,
\ee
for $p-q$ even, otherwise it is zero.

In order to check the correctness of this result and of all the used approximations, 
we compared Eq. (\ref{etaeta_TDL}) with the numerical data obtained directly from the starting expression (\ref{etaeta_int1}). 
We performed the numerical analysis fixing one of the modes (let us say $q$) and varying $p$ in an interval centred around $q$. 
We did such an analysis for several values of $q$ and, already for $N=50$, 
we obtained a very good agreement for $p+q$ sufficiently large.

\subsection{Time evolution of the  density profile}

Plugging the mode-mode correlator (\ref{etaeta_TDL}) into Eq. (\ref{nxtdef}), the time-dependent particle density reads
\be\label{nxt_1}
n(x,t) = 
\frac{2}{\pi L} \sum_{p,q \in D} \int_{0}^{\pi} du \,
\sin(p\pi x/L)\sin(q\pi x/L)
\frac{\cos[(u+\pi)(p-q)/2]}{1+\left[\frac{(p+q)\pi/(4N)}{1+\cos(u)}\right]^{2} }
{\rm e}^{i  \pi^2 t (p+q)(p-q)/L^2},
\ee
where the indices of the double sum run over the domain $D$ such 
that their difference $p-q$ is an even integer (i.e.  $p$ and $q$ are either both odd or both even). 
Therefore, the obvious change of variable is $p+q \equiv 2r$, $p-q\equiv 2l$; the domain $D$, in terms of these new variables,
becomes $1\leq r < \infty$, $-r+1 \leq l \leq r-1$. Then, Eq. (\ref{nxt_1})
can be written as
\be\label{nxt_2}
n(x,t) = 
\frac{1}{\pi L} \sum_{r =1}^{\infty}\sum_{l =-r+1}^{r-1} \int_{0}^{\pi} du \,
[\cos(2l\pi x/L)-\cos(2r\pi x/L)]
\frac{\cos[l(u+\pi)]}{1+\left[\frac{r\pi/(2N)}{1+\cos(u)}\right]^{2} }
{\rm e}^{i 4\pi^2 t r l /L^2}.
\ee
Since we are interested in the TDL, we introduce the rescaled variables $\tilde x = x/L$, $\tilde t = t/L$ and $\tilde r = r/L$. 
The sum over $r$ becomes an integral in the new variable $\tilde r$ and the sum over $l$ can be extended 
from $-\infty$ to $+\infty$, obtaining  
\bea%\label{nxt_3}
n( x, t) &=& 
\frac{1}{\pi} \int_{0}^{\infty} d\tilde r \sum_{l =-\infty}^{\infty} \int_{0}^{\pi} du \,
[\cos(2l\pi \tilde x)-\cos(2L\tilde r \pi \tilde x)]
\frac{\cos[l(u+\pi)]}{1+\left[\frac{\tilde r\pi/(2n)}{1+\cos(u)}\right]^{2} }
\cos( 4  \pi^2 \tilde t \tilde r l ) \nonumber \\
&\simeq&\frac{1}{2\pi} \int_{-\infty}^{\infty} d\tilde r \sum_{l =-\infty}^{\infty} \int_{0}^{\pi} du \,
\cos(2l\pi \tilde x)
\frac{\cos[l(u+\pi)]}{1+\left[\frac{\tilde r\pi/(2n)}{1+\cos(u)}\right]^{2} }
{\rm e}^{i 4  \pi^2 \tilde t \tilde r l },
\label{nxt_4}
\eea
where in the last line we dropped the term $\cos(2L\tilde r \pi \tilde x)$ because it is rapidly oscillating for $L\to\infty$.
In doing this approximation, we lose information about the behaviour close to the boundaries, but this is exactly what we forced 
when introducing the thermodynamic variable $\tilde x$.

%%%%%%% FIGURE n(x,t) %%%%%%%%
\begin{figure}[t]
 \includegraphics[width=0.5\textwidth]{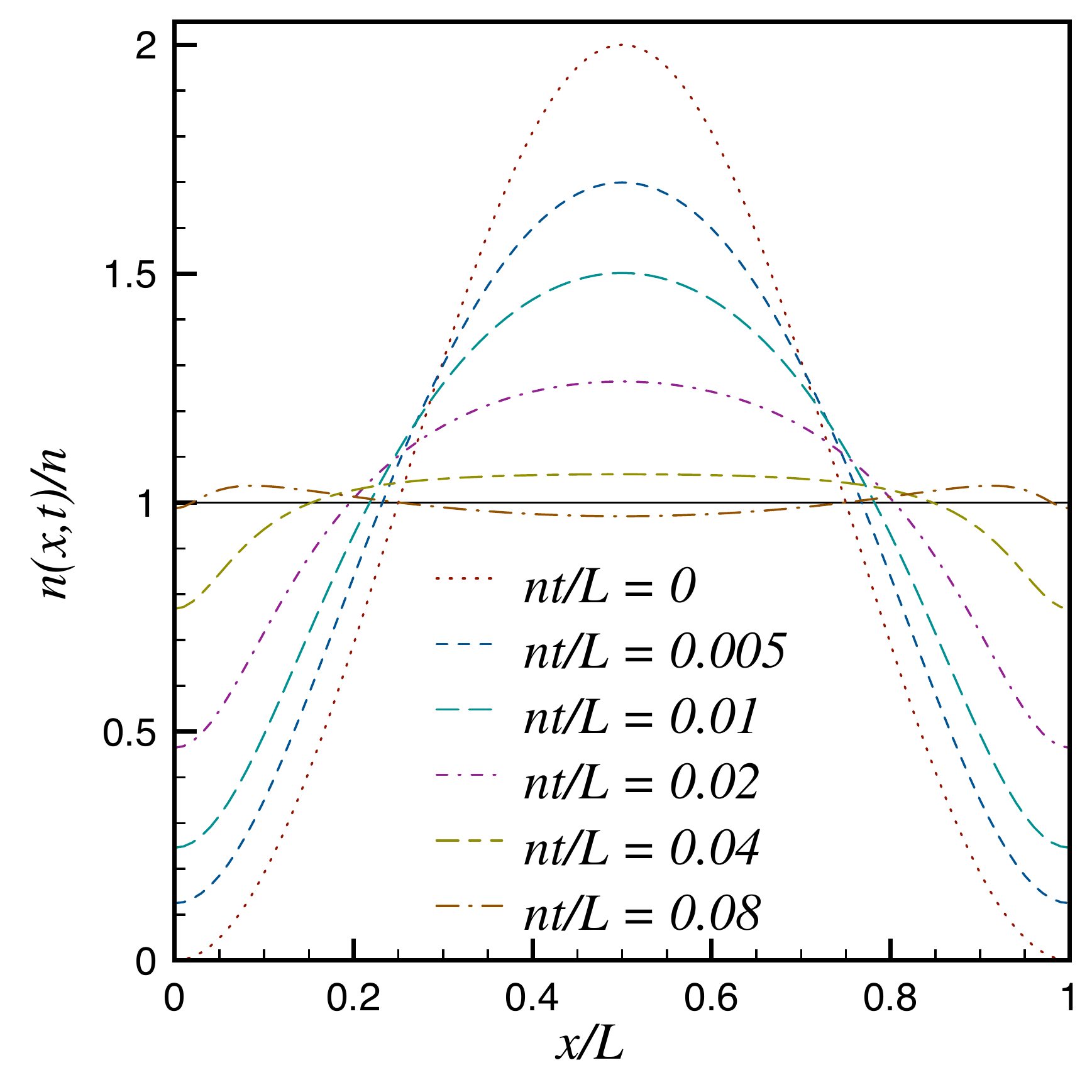}\includegraphics[width=0.5\textwidth]{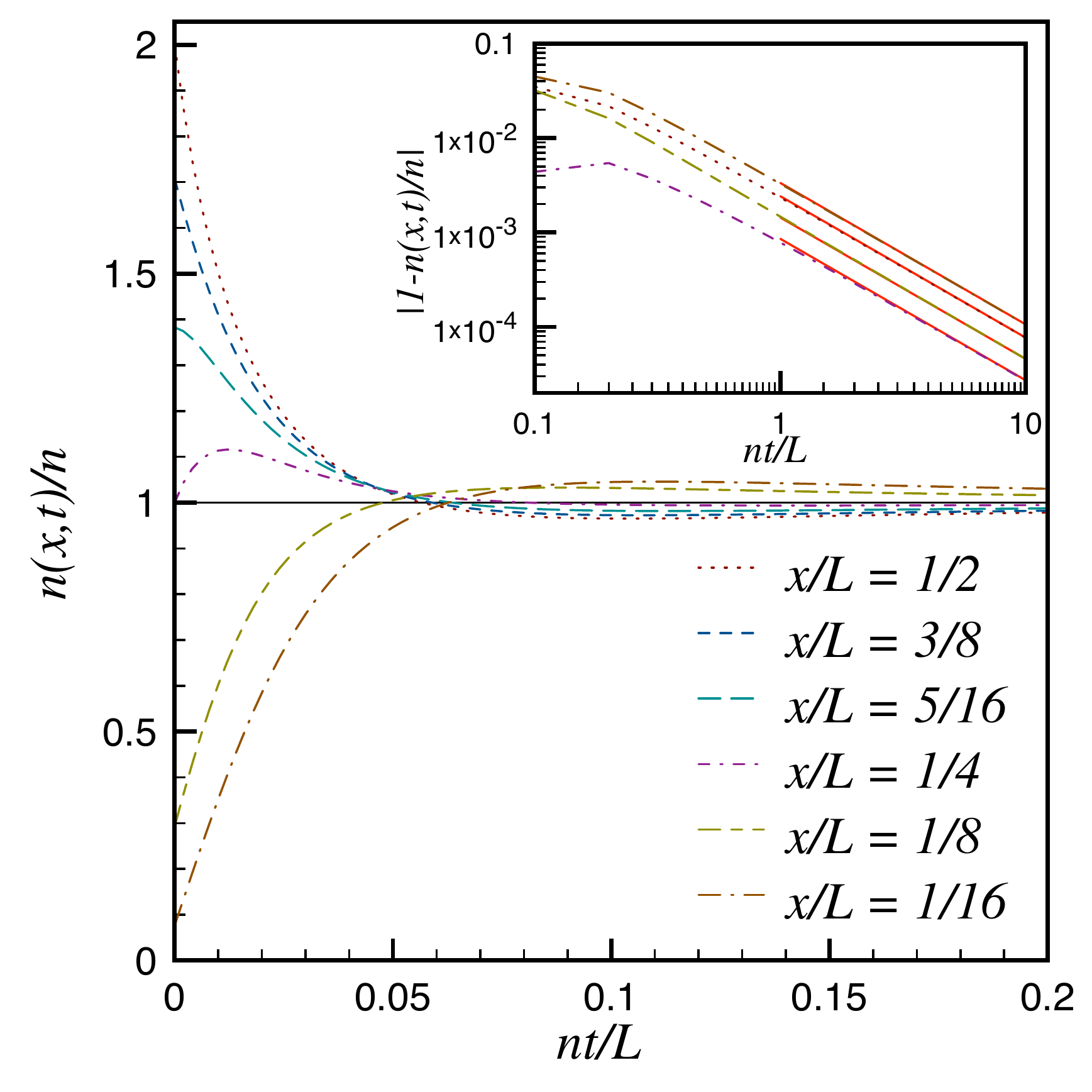}
\caption{(Left) 
The density profile $n(x,t)$ given by Eq. (\ref{nxt_TDL_v2}) as function 
of $\tilde x = x/L$ for fixed rescaled times $n\tilde t = n t/L$. 
The full black line is the uniform stationary value reached for $t\to\infty$.
(Right) The time evolution of the density $n(x, t)$ as function of the rescaled time  at fixed $\tilde x$. 
After a short transient, the density approaches the stationary value as a power-law (see the inset for a comparison, 
in $\log$-$\log$ scale, with the large-time behaviour given by Eq. (\ref{nxt_asym}) (red straight lines)).}
\label{fig5}
\end{figure}
 %%%%%%%%%%%%%%%%%%%%%%

The integral in $\tilde r$ in Eq. (\ref{nxt_4}) can be explicitly done giving
\bea\label{nxt_TDL}
n( x, t) 
& = & n \sum_{l =-\infty}^{\infty} (-1)^{l}\cos(2l\pi \tilde x) \int_{0}^{\pi} \frac{du}{\pi} \,
\cos(lu) [1+\cos(u)] 
{\rm e}^{-8 n \pi \tilde t | l | [1+\cos(u)]}\\
& = & n + 
2n \sum_{l =1}^{\infty} (-1)^{l}\cos(2l\pi \tilde x) \int_{0}^{\pi} \frac{du}{\pi} \,
\cos(lu) [1+\cos(u)] 
{\rm e}^{-8 n \pi  l \tilde t  [1+\cos(u)]}.\nonumber
\eea
%where we used the fact that $0 \leq 1+\cos(u)$ for $u\in[0,\pi]$.
This result does not depend on $L$ and $N$ independently, but only on their ratio $n=N/L$, as it should.

The just derived $n(x,t)$ correctly reproduces the two limiting cases $t=0$ and $t\to\infty$. 
The former is obtained by setting $\tilde t=0$ in Eq. (\ref{nxt_TDL}) and then the integral can be easily evaluated, 
giving for the density
\be
n( x, 0) = 
 n \sum_{l =-\infty}^{\infty} \frac{(-1)^{l}\sin[\pi l]}{\pi l(1-l^{2})} \cos(2l\pi \tilde x)
 =  2 n \sin^{2} (\pi \tilde x),
\ee
where we used the fact that the only non-zero contributions to the sum
come from $l=0$ and $|l|=1$.
The limit $t\to\infty$ is  given  only by the $l=0$ term in the sum 
(since all the others are exponentially suppressed), and it agrees with the bulk  stationary result $n$. 

More explicit information can be extracted from Eq. (\ref{nxt_TDL}) using the 
integral representation of the modified Bessel functions (valid for $m\in \mathbb{Z}$ and $s\in\mathbb{R}$)
\be
{\rm I}_{m}(s) = (-1)^{m}  \int_{0}^{\pi} \frac{du}\pi \, \cos(m u) e^{-s\cos(u)} ,
\ee
and the identity 
$\p_s {\rm I}_{m}(s) = [ {\rm I}_{m-1}(s)+ {\rm I}_{m+1}(s)]/2$. 
Therefore, Eq. (\ref{nxt_TDL}) can be written as
\bea
n( x, t) 
& = & n - 2n \sum_{l=1}^{\infty} \cos(2l\pi \tilde x) \,
\p_s\left[{\rm I}_{l}(s) {\rm e}^{-s}\right]  \Big|_{s = 8n\pi l \tilde t} \nonumber \\
& = & n + n \sum_{l=1}^{\infty} \cos(2l\pi \tilde x) \,
{\rm e}^{-8n\pi l \tilde t}
\left[ 2{\rm I}_{l}(8n\pi l \tilde t)-{\rm I}_{l-1}(8n\pi l \tilde t)-{\rm I}_{1+l}(8n\pi l \tilde t) \right] .
\label{nxt_TDL_v2}
\eea
This form shows cleanly how the time-dependent density approaches the  stationary value. 
Indeed, even if the presence of the exponential factor could suggest a typical relaxation time, 
the combination with the Bessel functions gives rise to an algebraic decay for large times. 
Indeed, the relaxation of the density takes place in a two-step process. 
First there is a short transient for $n \tilde t  \ll 1$, in which the density decays very quickly to a value very close to the stationary 
one, see Fig. \ref{fig5}. After this transient, the relaxation gets dramatically slowed down to an algebraic behaviour. 
Indeed, the use the asymptotic expansion
$\left[ 2{\rm I}_{l}(s)-{\rm I}_{l-1}(s)-{\rm I}_{1+l}(s) \right] \exp(-s) \sim s^{-3/2}/\sqrt{2\pi}$ leads us
to the following large-time behaviour of the  density (see Fig. \ref{fig5})
\be\label{nxt_asym}
n( x,  t) \sim n + \frac{n}{64\pi^2 (n \tilde t)^{3/2}} 
\left[ {\rm Li}_{3/2}({\rm e}^{2\pi i \tilde x}) +{\rm Li}_{3/2}({\rm e}^{-2\pi i \tilde x}) \right] 
\quad {\rm for} \quad \tilde t \gg 1,
\ee
in terms of the Polylogarithm function ${\rm Li}_{m}(s) \equiv \sum_{k=1}^{\infty} s^{k}/k^{m}$.
 
The physical interpretation of this two-step relaxation behaviour is very intuitive. 
Indeed, soon after the quench, the bosons experience an infinite strong repulsion 
which suddenly tends to reduce the density in the centre by moving the particles close to the boundaries.
However, after this quick process the final equilibration takes place by means of a series of many bounces off the 
boundaries and this process needs times which are much larger than $L/n$ (we recall that the speed of sound in the 
Tonks Girardeau gas and in our normalisation is $v=2\pi n$).

\subsection{Time evolution of the two-point fermionic correlation function}

The time-dependent two-points fermionic correlation function $C(x,y;t)$ can be evaluated along the same lines as the density. 
Plugging the mode-mode correlator (\ref{etaeta_TDL}) into Eq. (\ref{fermi_corr_t_eta}), we have
\be\label{Cxt_1}
C(x,y;t) = 
\frac{2}{\pi L} \sum_{p,q \in D} \int_{0}^{\pi} du \,
\sin\Big[ \frac{p\pi (w+z)}{2L} \Big] \sin \Big[ \frac{q\pi (w-z) }{ 2L} \Big]
\frac{\cos[(u+\pi)(p-q)/2]}{1+\left[\frac{(p+q)\pi/(4N)}{1+\cos(u)}\right]^{2} }
{\rm e}^{i t \pi^2 (p+q)(p-q)/L^2},
\ee
where we introduced the variables $z=x-y$, $w=x+y$. 
As shown in Appendix \ref{appC}, as long as we are interested to
the TDL of this correlator in the bulk, we can perform the following replacement in the integral
$
\sin\left[ p\pi (w+z) / (2L) \right] \sin \left[ q\pi (w-z) / (2L) \right] 
\to \cos\left[ (p+q)\pi z / (2L) \right] \cos \left[ (p-q) \pi w / (2L) \right] /2.
$

%%%%%%% FIGURE C0 and C(x,y;t) %%%%%%%%
\begin{figure}[t]
\includegraphics[width=0.6\textwidth]{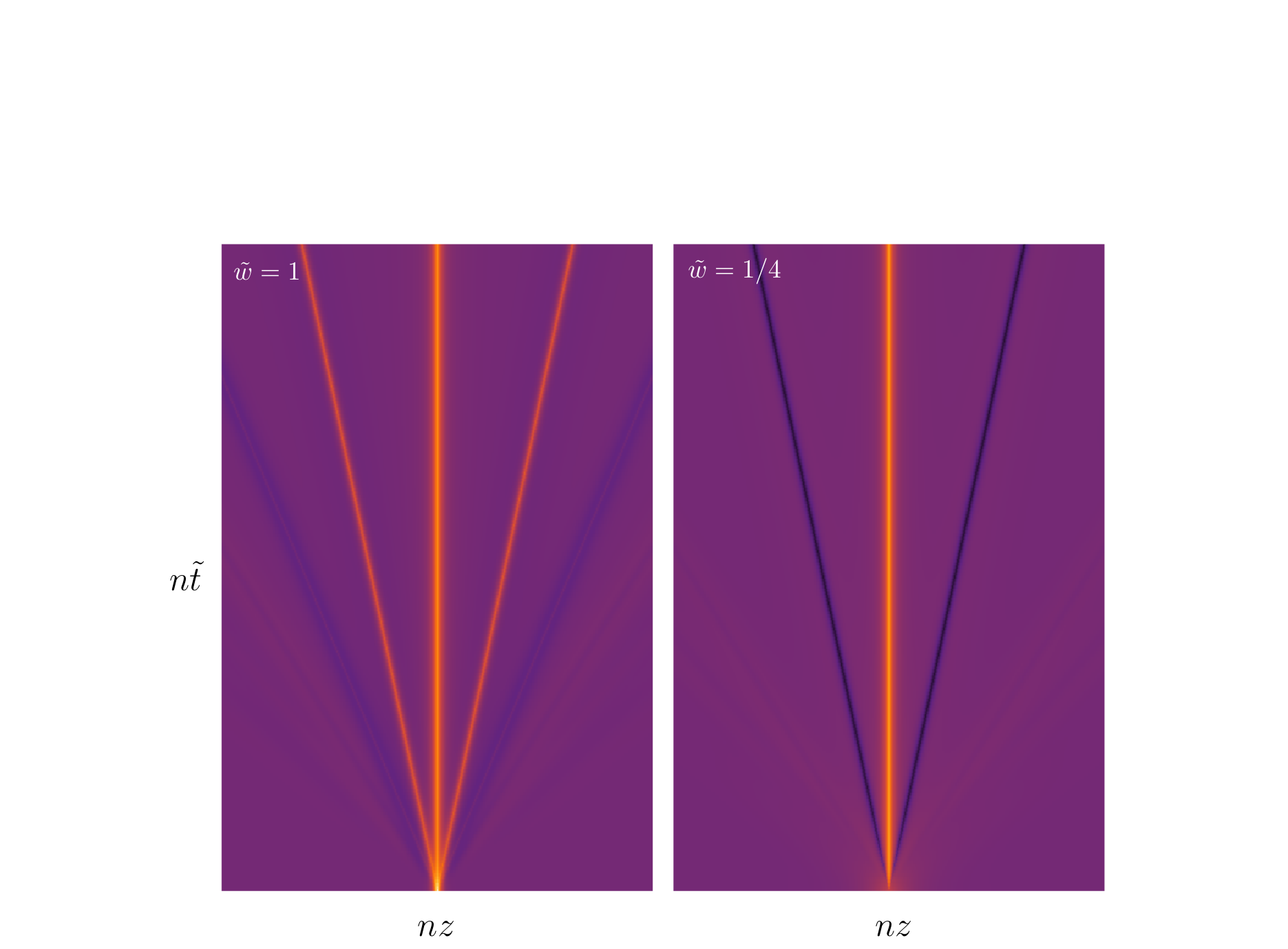}
\includegraphics[width=0.5\textwidth]{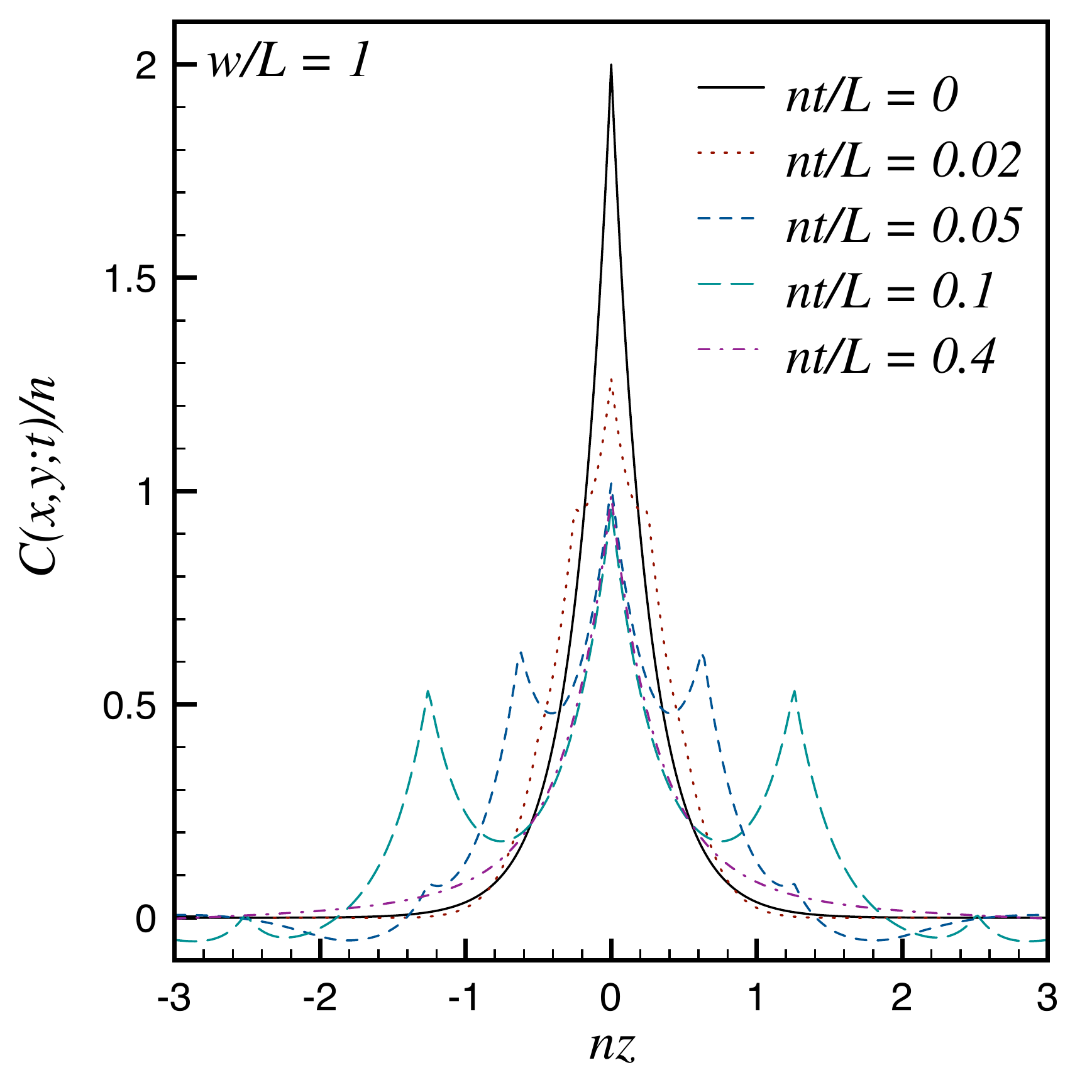}\includegraphics[width=0.5\textwidth]{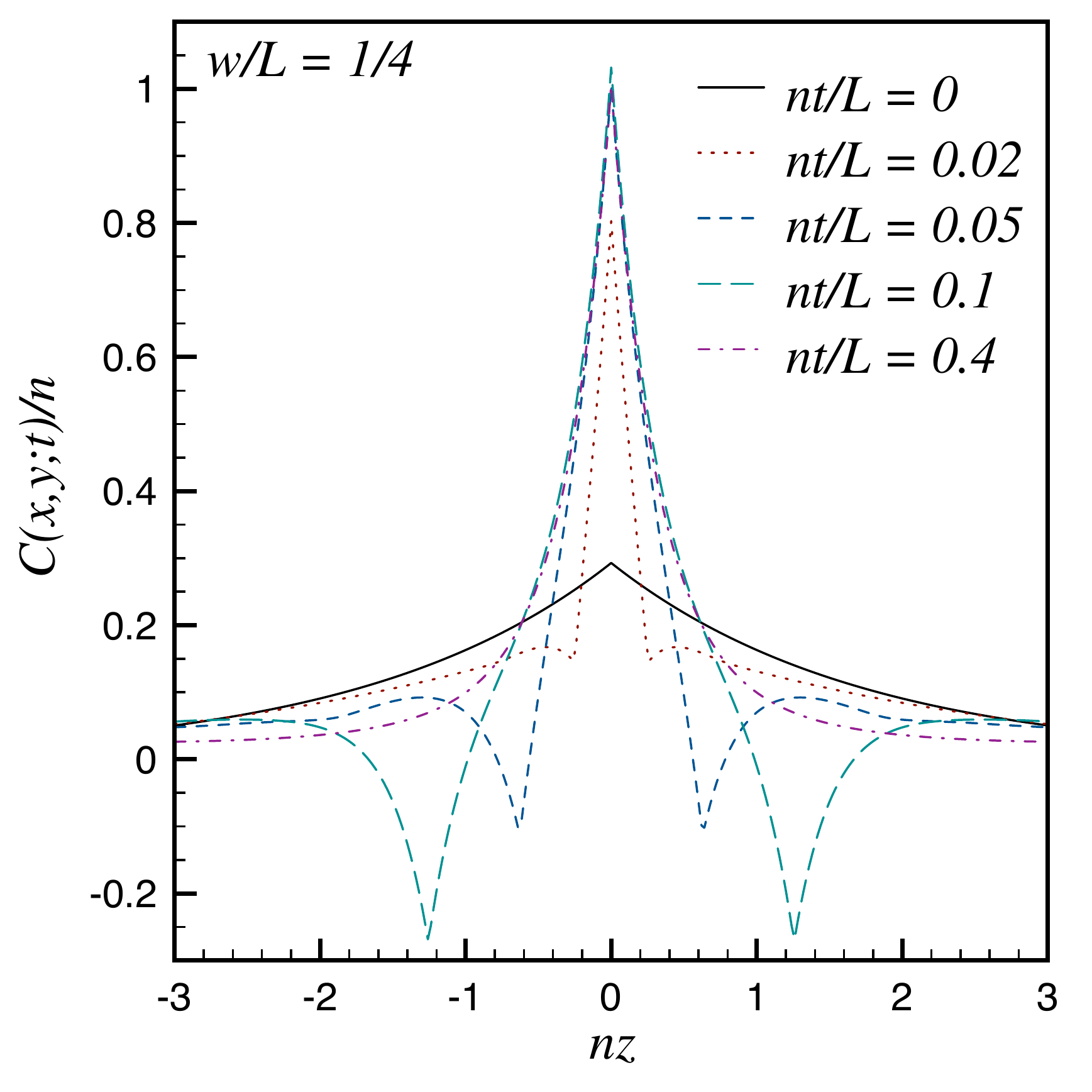}
\caption{
Top: Space-time contour plot of the time dependent fermionic correlator $C(x,y;t)$ at fixed $\tilde w\equiv w/L=(x+y)/L$.
The time is rescaled as $\tilde t = t/L$ and $z=x-y$. The plotted region corresponds to $ n \tilde t  \in [0,1]$ e $n z \in [-20,20]$.
Notice the correlation peaks expelled from $z=0$ which move ballistically with velocities that are integer multiples of 
$v_{\rm p}=4\pi/L$.
Bottom:
Profiles of the fermionic correlation function as function of 
$z$ for different rescaled time $n\tilde t = nt/L$ and $w/L = 1,\,1/4$
(i.e. each curve is a horizontal cut of the contour plot on the top).}
\label{fig6}
\end{figure}
 %%%%%%%%%%%%%%%%%%%%%%

Then, changing the indices of the sum as $p+q = 2r$ and $p-q=2l$,  the correlator can be written as  
\be\label{Cxt_2}
C(x,y;t) = 
\frac{1}{\pi L} \sum_{r=1}^{\infty} \sum_{l=-r+1}^{r-1} \int_{0}^{\pi} du \,
\cos( r \pi z / L) \cos ( l \pi w / L )
\frac{\cos[l(u+\pi)]}{1+\left[\frac{r\pi/(2N)}{1+\cos(u)}\right]^{2} }
{\rm e}^{i 4 t \pi^2 r l / L^2}.
\ee
In analogy with the density profile, we introduce the rescaled variables
\be
\tilde r = r/L, \quad \tilde t = t/L, \quad {\rm and} \quad \tilde w = w/L, 
\label{rescale}
\ee
but we do not rescale the distance between the two points $z=x-y$, 
in such a way to explore correlations at arbitrary distances in the bulk.  
We replace the sum over $r$ with an integral over $\tilde r$, we let the sum over $l$ to run from $-\infty$
to $\infty$, and we use simple trigonometric identities to  write the correlator as 
\be\label{Cxt_3}
C(x,y;t) = 
\frac{1}{4 \pi} \int_{-\infty}^{\infty} d\tilde r \sum_{l=-\infty}^{\infty} \int_{0}^{\pi} du \,
 \cos ( l \pi \tilde w )
\frac{\cos[l(u+\pi)]}{1+\left[\frac{\tilde r\pi/(2n)}{1+\cos(u)}\right]^{2} }
\left[\cos( \tilde r \pi z + 4\pi^2 \tilde r \tilde t l) +\cos( \tilde r \pi z -4\pi^2 \tilde r \tilde t l)\right].
\ee
Once again, the integral in $\tilde r$ can be easily done, being proportional to the Fourier 
transform of a Lorentzian function, obtaining
\bea%\label{Cxt_4}
C(x,y;t) &=& 
n \sum_{l=-\infty}^{\infty} 
(-1)^{l} \cos ( l \pi \tilde w )
\int_{0}^{\pi} \frac{du}{2\pi} \,
\cos(lu)[1+\cos(u)]
\left\{{\rm e}^{-2n|z+4\pi l \tilde t|[1+\cos(u)]}+{\rm e}^{-2n|z-4\pi l \tilde t|[1+\cos(u)]} \right\},\nonumber\\
\label{Cxt_5}
&=&  -\frac{n}{2} \sum_{l=-\infty}^{\infty} 
\cos ( l \pi \tilde w )
\left\{ \p_s \left[{\rm I}_{l}(s) {\rm e}^{-s}\right] \Big|_{s=2n|z+4\pi l \tilde t |}
+ \p_s \left[{\rm I}_{l}(s) {\rm e}^{-s}\right] \Big|_{s=2n|z-4\pi l \tilde t |} \right\}.
\eea
We can now isolate the $l=0$ term, which corresponds to the bulk stationary result 
$C_{\infty}^B(z)$ in Eq. (\ref{CGGE_TI})), and we can rewrite Eq. (\ref{Cxt_5}) as
 \be\label{Cxt_6}
C(x,y;t) = C_{\infty}^B(z)
-n\sum_{l=1}^{\infty} 
\cos ( l \pi \tilde w )
\left\{ \p_s \left[{\rm I}_{l}(s) {\rm e}^{-s}\right] \Big|_{s=2n|z+4\pi l \tilde t |}
+ \p_s \left[{\rm I}_{l}(s) {\rm e}^{-s}\right] \Big|_{s=2n|z-4\pi l \tilde t |} \right\}.
\ee

Let us now critically analyse this time-dependent correlation function. 
For finite rescaled time $\tilde t < \infty$, Eq. (\ref{Cxt_6}) is not translational invariant since it depends both on $z$ and $\tilde w$.
As $\tilde t\to\infty$, all terms with $l\neq 0$ in the sum vanish and only the translational invariant 
stationary part survives.
In the opposite limit $\tilde t=0$, Eq. (\ref{Cxt_6}) should reproduce the scaling regime of 
the initial correlation fuction $\langle \hat\Psi^{\dag}(x)\hat\Psi(y)\rangle$  in Eq. (\ref{PsiPsi_0_TDL}).
This is indeed not so apparent from the series representation of the correlation function, but can be shown 
plugging the following  infinite sum 
\be
\sum_{l=-\infty}^{\infty} a^{l} {\rm I}_{l}(s) = \exp [s(a+1/a)/2],
\ee  
into Eq. (\ref{Cxt_5}), obtaining 
\bea
 C(x,y;0) & = & -n \, \p_s \left[ {\rm e}^{-s}\, \sum_{l=-\infty}^{\infty}
 {\rm Re} \left({\rm e}^{i l\pi \tilde w}\right) \, {\rm I}_{l}(s)\right]_{s=2n|z|}
 \nonumber \\ & = & 
  -n \, \p_s  \left.  {\rm e}^{-[1-\cos(\pi \tilde w)] s }  \right|_{s=2n|z|}
  =  n\,[1-\cos(\pi\tilde w)] e^{ -2 n [1-\cos(\pi \tilde w)] |z| },
  \label{C0_scaling}
\eea
which coincides with Eq. (\ref{PsiPsi_0_TDL}).

%%%%%%% FIGURE C(x,y;t) vs w,z,t %%%%%%%%
\begin{figure}[t]
\includegraphics[width=0.8\textwidth]{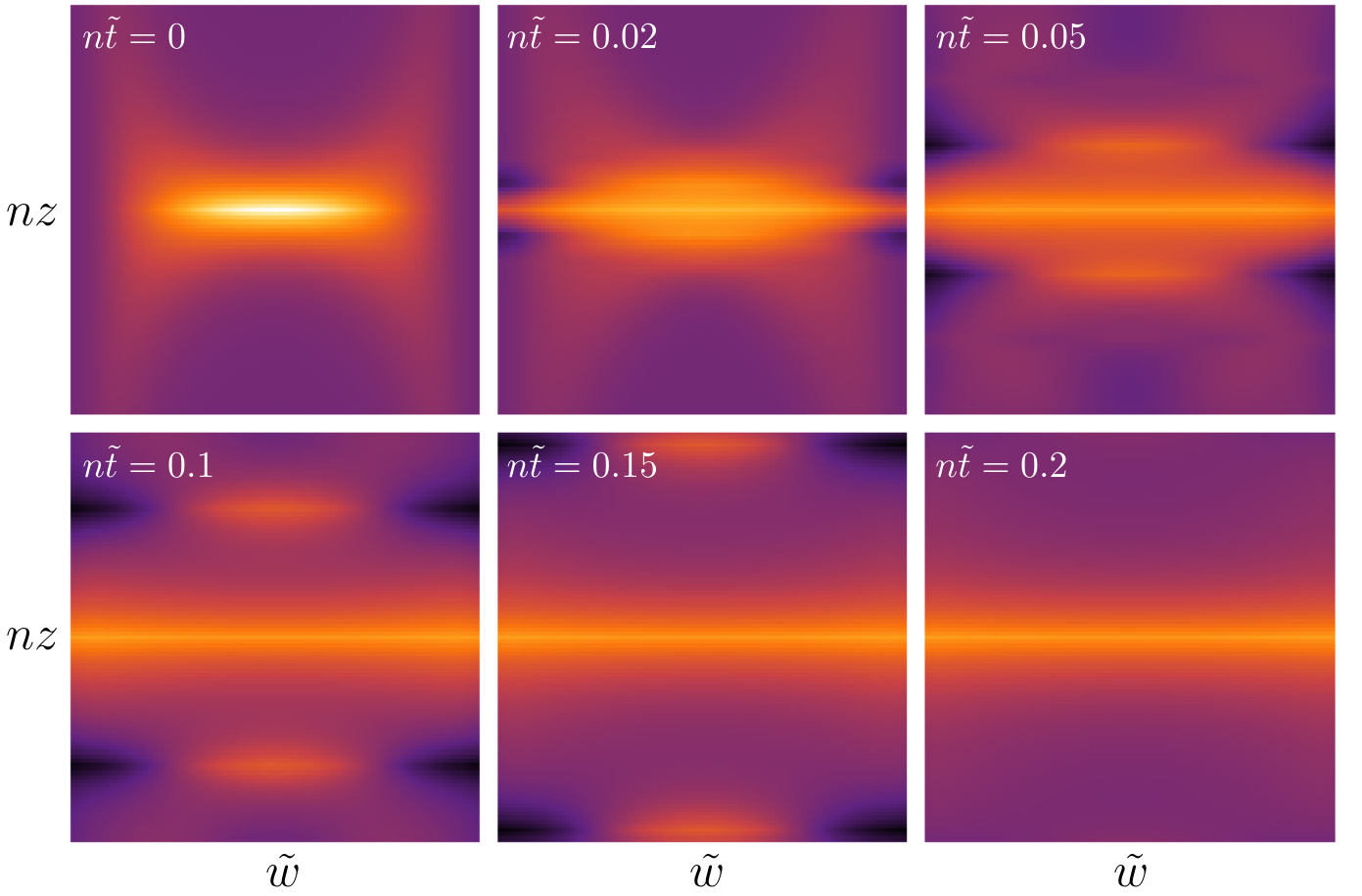}
\caption{Color snapshots of the fermionic correlation $C(x,y;t)$ given by Eq. (\ref{Cxt_6})
for different rescaled times $n\tilde t = n t/L$ as a function of $nz=n(x-y)\in[-2,2]$ 
and $\tilde w =(x+y)/L \in [0,2]$. Notice the initial strong inhomogeneity along $\tilde w$ 
which is smoothed out during the time evolution.}
\label{fig7}
\end{figure}
 %%%%%%%%%%%%%%%%%%%%%% 
 
The time-evolved correlation function in Eq. (\ref{Cxt_6}) depends both on $z$ and $\tilde w$.
These two variables work on different scales: 
(i) $z = x-y$  is a ``local'' variable and  indeed, it is the only one that survives  in the stationary state; 
(ii) $\tilde w$ is a global variable on which $z$ is modulated. 
Therefore, in order to understand the physics of Eq. (\ref{Cxt_6}) for different times, it is useful to fix the value of
$\tilde w$ and plot the the time-dependent correlator as a function of the local variable $z$ as 
done in Fig. \ref{fig6}.
The resulting behaviour is reminiscent of the one observed in other quench studies with inhomogeneous initial states \cite{csc13,ck14}. 
Indeed, the correlation function approaches the stationary value by expelling a series of traveling peaks from the vicinity of $z\simeq0$
which afterwards move ballistically through the system. 
The velocity of the primary peaks (i.e. the highest and most visible ones in Fig. \ref{fig6}) is $v_{\rm p}=4\pi/L$.
There is an infinite number of smaller secondary peaks (a second one is visible at a close look of Fig. \ref{fig6}) 
which move with velocities which are integer multiples of $v_{\rm p}$.
This aspect is independent from the precise value of $\tilde w$, while the other details of this process (e.g. shape and amplitude 
of the peaks)  depend on the value of $\tilde w$ (as it should be clear from Fig. \ref{fig6}). 
Consequently, the characteristic time in which the correlator at distance $z$ gets close to its stationary value is almost independent 
from $\tilde w$, because it is roughly the time needed for the primary peaks to travel a distance $z$.

To conclude our analysis, in Fig. \ref{fig7} we report a contour plot of $C(x,y;t)$ as a function of $nz$ and $\tilde w$ for different times. 
These plots pictorially show how the initial inhomogeneous correlation is smoothed out and made uniform 
by counter propagating fronts emitted from $z=0$.
Consequently, there is an effective region inside a horizon $|z|< v_{\rm p} t$ in which the system is almost stationary 
and translational invariant (and hence approximately described by $C_{\infty}^B(z)$).
Note that the horizon introduced above to explain the equilibration of the correlation function is very different from the standard 
picture for translational invariant and homogeneous quenches in which the horizon is governed by the velocity $v$ of the elementary 
quasi-particle excitations produced in pairs of opposite momentum  \cite{cc-06,fm-10,sfm-12,CEFI,cetal-12}. 
In the present case (in analogy to Refs. \cite{csc13,ck14}) $v_{\rm p}\ll v=2\pi n$ and the equilibration is much slower 
than in a homogeneous system.

Finally it is worth stressing that, as an important difference with the stationary state, the two-point fermionic correlation
at finite time does not univocally characterise the time-dependent state, because Wick's theorem is not valid and 
multi-point correlations must be calculated on a case by case basis.
In particular, we cannot write the bosonic two-point function as a Fredholm's minor for finite times. 
 
\section{Conclusion}\label{sec5}

We analysed the effect of a hard-wall trapping potential in the one-dimensional Bose gas following a quantum quench from  
free to hard-core bosons. 
Both the initial condition and the Hamiltonian governing the time evolution break translational invariance.
As a consequence, the density and the two-point correlation function exhibit a nontrivial space-time dependence.
Although this is a quench between two free theories, the pre- and post-quench mode-operators are not linearly related and, 
therefore, the time evolution shows many non-trivial effects like the breaking of Wick's theorem for finite times.

We studied in detail both the large time behaviour and the full time evolution of the 
density profile and of the two-point fermionic correlation function.
The large time properties turned out to be described by the GGE constructed with the mode 
occupation numbers and hence Wick's theorem is restored for large time. 
Although the system is not translational invariant, the stationary density is uniform.
The bulk correlation function turned out to depend only on the distance between 
the two points and so we conclude that translational invariance is dynamically restored 
(in the GGE Wick's theorem applies and so all correlations can be derived from the two-point one) apart 
from finite size effects close to the boundaries.
However, the stationary state keeps memory of the initial inhomogeneous state because 
the asymptotic two-point fermionic correlation function decays algebraically for large (bulk) distances
while in the periodic case the decay has been found to be always exponential \cite{kcc14}. 
We point out that this very peculiar effect is mainly due to the highly inhomogeneous initial state, 
which is a consequence of the bosonic nature of the initial state in which all the particles are in the 
same one-particle state.
This does not happen for quenches in ``purely fermionic'' theories (such as the Ising chain studied in Refs. \cite{ir-10}) 
for which a hard-wall trapping potential leads to a less inhomogeneous initial state because of
the effective repulsion due to the Pauli principle.    
To complete the analysis of the stationary state, we also computed numerically the bosonic correlation function
which turned out to decay exponentially for large distances. 
 The decay rate (i.e. the correlation length) is however different from the one in the PBC case.

We also studied the full time-dependence for both the particle density and the
two-point fermionic correlation function. 
We found that the relaxation takes place with a two-step mechanism:
 first there is a rather quick transient in which the density drops to an almost uniform value, 
and after the decay to the stationary value is algebraic and it is driven by the particles bouncing off the boundaries 
many times. 
The equilibration of the two-point function takes place through the expulsion of a series of correlation peaks that move out ballistically 
leaving the system almost equilibrated inside an effective horizon. 

It would be extremely interesting to generalise our findings to the case of the experimentally more relevant 
harmonic trapping potential. 
However, the algebra becomes immediately very cumbersome because the one-particle eigenfunctions
are the Hermite polynomials and not simple trigonometric functions. 
In light of the results we found for the hard-wall confinement, it seems very unlikely that an approach based on local 
density approximation could provide the correct answer, making an exact calculation for the harmonic trap even more desirable. 

Another, more difficult, generalisation would be to consider the same quench in the presence of a hard-wall trap but to a 
finite interaction Lieb-Liniger model (which is integrable \cite{hwint}). 
However, the overlaps needed in the Bethe ansatz framework \cite{ce-13} are very difficult to 
calculate compared to the periodic case in which the initial many-body wave-function is constant \cite{nwbc-13,b-14}.

\section*{Acknowledgments}   
PC and MC acknowledge the ERC  for financial  support under Starting Grant 279391 EDEQS. 
MK acknowledges financial support from the Marie Curie IIF Grant PIIF-GA-2012- 330076.

\appendix

%%%%%%%%%%%%%%% LATTICE FORMULATION %%%%%%%%%%%%%%%%

\section{Lattice formulation}\label{appA}

In this appendix we provide a rigorous lattice regularisation to justify the identity (\ref{HCtoBoson}).
We closely follow the analogous calculation for PBC \cite{kcc14}.

Let us consider a system of $N$ bosons hopping on a one-dimensional lattice 
with $M$ sites with lattice spacing $\delta$; the length of the lattice is $L=M\delta$.
The one-particle eigenfunction associated
to the lowest energy level is $\sqrt{2/M}\sin(\pi i /M)$, and so the many-body ground state is given by 
 \be\label{initial_state}
|N\rangle = \sqrt{\frac{2^N}{M^NN!}}\left(\sum_{i=1}^{M} 
\sin\left(\frac{\pi i}{M}\right)\hat b^{\dag}_{i}\right)^N|0\rangle,
\ee
where $\hat b^{(\dag)}_i$ are the canonical bosonic operators acting on the $i^{th}$ site and
$|0\rangle = \prod_{i}|0\rangle_i$ with $|n\rangle_i$ being the $n$-boson state at site $i$.

The hard-core boson operators are defined as in the continuum case
\be\label{HC}
\hat a_i=P_i \hat b_i\hat P_i, \quad \hat a^{\dag}_i=P_i \hat b^{\dag}_i\hat P_i,
\ee
with $P_i=|0\rangle\langle 0|_i+|1\rangle\langle1|_i$ being the on-site projector
on the truncated Hilbert space. The hard-core boson operators satisfy the {\it mixed} algebra
\bea
[\hat a_i,\hat a_j]=[\hat a^{\dag}_i,\hat a^{\dag}_i]=[\hat a_i,\hat a^{\dag}_j]=0,& \quad & i\neq j,\nonumber \\
\hat a^{2}_i = \hat a^{\dag 2}_i = 0, &\quad & \{\hat a_i,\hat a^{\dag}_i\} = 1.
\eea

The Jordan-Wigner mapping from hard-core bosons to free fermions on the lattice reads
\be
\hat a_i=e^{-i\pi\sum_{j<i}\hat c^{\dag}_j \hat c_j} \hat c_i
=\prod_{j<i}(1-2\hat c^{\dag}_{j}\hat c_j)\hat c_i, \quad
\hat c_i=e^{i\pi\sum_{j<i}\hat \hat a^{\dag}_j \hat a_j}\hat a_i
=\prod_{j<i}(1-2\hat a^{\dag}_{j}\hat a_j) \hat a_i,
\ee
with $\{\hat c_i,\hat c^{\dag}_i\}=\delta_{i,j}$. 
The lattice thermodynamic limit is defined as $N,M\to\infty$, 
keeping the filling factor $\nu = N/M$ constant.
The continuum limit for finite systems is obtained by considering
the lattice spacing $\delta\to 0$, the number of sites $M\to\infty$,
while the physical length $L=M\delta$ is kept constant. 
Therefore, the continuum TDL can now be taken as $N,L\to\infty$, with gas density $n=N/L$ constant 
(we can equivalently think to the continuum and thermodynamic limit as the limit $\delta,\nu\to 0$, keeping constant $n=\nu\delta$).
Finally, the relations between the lattice  and the continuum operators are
\be
\hat b_{i}=\sqrt{\delta}\hat \phi(\delta i),\quad
\hat a_{i}=\sqrt{\delta}\hat\Phi(\delta i),\quad
\hat c_{i}=\sqrt{\delta}\hat\Psi(\delta i).\quad
\ee

The initial fermionic correlation function for $k<l$ can be written as
\bea
\langle N|\hat c^{\dag}_{k} \hat c_{l}|N\rangle
& = & \langle N| \hat a^{\dag}_{k}\prod_{j=k+1}^{l-1}(1-2\hat a^{\dag}_j \hat a_j) \hat a_{l}|N\rangle \nonumber  \\
& = & \sum_{r=0}(-2)^{r} \sum_{k<n_1<\ldots<n_r<l} 
\langle N|\hat a^{\dag}_{k} \hat a^{\dag}_{n_1} \hat a_{n_1} 
\dots \hat a^{\dag}_{n_r} \hat a_{n_r} \hat a_{l} |N\rangle.
\label{corfun1}
\eea
Therefore, in order to find the fermionic correlation function 
we have to evaluate the multipoint hard-core boson correlators
\be
\langle N|\hat a^{\dag}_{k}\hat a^{\dag}_{n_1}a_{n_1}\dots \hat a^{\dag}_{n_r}\hat a_{n_r}\hat a_{l}|N\rangle,
\label{string1}
\ee
which can be calculated by expand the multinomial in Eq. (\ref{initial_state}) as
\be
|N\rangle=\sqrt{\frac{2^N}{M^N N!}}\sum_{i_1,\ldots,i_M}\binom{N}{i_1,\ldots,i_M}
(p_1\hat b^{\dag}_1)^{i_1}\dots(p_M \hat b^{\dag}_M)^{i_M}|0\rangle,
\label{initial_state1}
\ee
where $p_i \equiv \sin(\pi i/M)$ and the sum runs over all sets of non-negative integers
$\{i_1,\ldots,i_M\}$ such that $\sum_{j} i_j = N$.
Let us start by considering the action of the hard-core boson string in Eq. (\ref{string1})
on the many-body ground state, i.e.  
$
\hat a^{\dag}_{k}\hat a^{\dag}_{n_1}\hat a_{n_1}\dots \hat a^{\dag}_{n_r}\hat a_{n_r}\hat a_{l}|N\rangle.
$
If we want a non-zero result,  we must fix the value of some indices, i.e. $i_l=i_{n_r}=\dots i_{n_1}=1$ and $i_k=0$. 
This comes from having rewritten the hard-core boson operators $\hat a^{(\dag)}$ 
in terms of the canonical ones $\hat b^{(\dag)}$: the projectors $P_i$ appear and they annihilate 
all the multi-occupied sites, thus obtaining
\bea\label{ket1}
\hat a^{\dag}_{k}\hat a^{\dag}_{n_1}\hat a_{n_1}\dots \hat a^{\dag}_{n_r}\hat a_{n_r}\hat a_{l}|N\rangle
& = &
\sqrt{\frac{2^N}{M^N N!}}\sum_{\{i_1,\ldots i_M\}'}
\binom{N}{i_1,\ldots, i_k=0,\ldots,i_{n_1}=1,\ldots,i_{n_r}=1,i_l=1,\ldots i_M}\nonumber \\
&& \times 
(p_1b^{\dag}_1)^{i_1}\dots(p_Mb^{\dag}_M)^{i_M}|0\rangle,
\eea
wherein $\{i_1,\ldots i_M\}'=\{i_1\dots i_M\} \backslash \{i_k,i_{n_1},\dots i_{n_r}, i_l\}$. Notice that in the
previous equation all $\hat b^{\dag}_{n_j}$ and $\hat b^\dag_k$ come with power one, while there is no
$\hat b^\dag_l$. 
When we consider the scalar product between the state defined in Eq. (\ref{ket1}) and the 
ground state $|N\rangle$ the only non-zero contributions  come from those terms which perfectly
match the powers of all operators. Therefore, by using
$
\langle0|(p_i \hat b_i)^n (p_i \hat b^{\dag }_i)^n|0\rangle=p_i^{2n}n!
$
we obtain 
\be\label{scalar_product}
\langle N|\hat a^{\dag}_{k}\hat a^{\dag}_{n_1}a_{n_1}\dots 
\hat a^{\dag}_{n_r}\hat a_{n_r}\hat a_{l}|N\rangle
=\frac{2^N}{M^NN!}p_k p_l \prod_{j=1}^{r}p^2_{n_j}
\sum_{\{i_1,\ldots i_{M}\}'}\binom{N}{i_1,\dots, i_{M}}^2p_{1}^{2i_1}i_{1}!
\dots p_{M}^{2i_{M}}i_{M}!,
\ee
which, since $\sum_{j\,:\,i_j\in\{i_1,\ldots i_{M}\}'}i_j=N-r-1$, can be rewritten as
\be\label{result}
\langle N|\hat a^{\dag}_{k}\hat a^{\dag}_{n_1}\hat a_{n_1}\dots 
\hat a^{\dag}_{n_r}\hat a_{n_r}\hat a_{l}|N\rangle
=\left(\frac{2}{M}\right)^{N}p_kp_l\,p_{n_1}^2\dots p_{n_r}^2N(N-1)\dots(N-r)
\left(\sum_{j \, : \, i_j\in\{i_1,\ldots i_{M}\}'}p^2_j\right)^{N-r-1},
\ee
where, once again, we used the definition of the multinomial expansion.
The indices $\{i_1,\ldots i_{M}\}'$ are $M-r-2$ and their distribution depends on 
how the other $r+2$ indices, namely $k,l,n_1,\ldots,n_r$, have been chosen on the lattice. 
In the continuum limit between the site $l$ and $k$ there is an infinite number of operators, 
however $r\leq N-1$ since the string $\hat a^\dag_{n_1} \hat a_{n_1}\dots \hat a^\dag_{n_r} \hat a_{n_r}$
acts on $(N-1)$-particle state with $N$ finite. Moreover, in such a limit, the lattice holds
an infinitely dense number of sites and therefore the index 
$j$ such that  $i_j\in \{i_1,\ldots i_{M}\}' $ runs over the whole lattice except for  $r+2$
positions which represent a subset of null measure in the continuum limit. Thus the following 
approximation holds 
\be
\sum_{j \, : \, i_j\in\{i_1,\ldots i_{M}\}'}p^2_j 
\simeq \frac{1}{\delta} \int_{0}^{L} dz \, \sin^{2}\left(\frac{\pi z}{L}\right) = \frac{M}{2},
\ee
which leads to
\be\label{result1}
\langle N|\hat a^{\dag}_{k}\hat a^{\dag}_{n_1}\hat a_{n_1}\dots 
\hat a^{\dag}_{n_r}\hat a_{n_r}\hat a_{l}|N\rangle
\simeq \left(\frac{2}{M}\right)^{r+1}N(N-1)\dots(N-r) p_kp_l\,p_{n_1}^2\dots p_{n_r}^2.
\ee
At this point, in order to calculate the fermionic two-point function, we have to sum 
terms like those in Eq. (\ref{result}) over the indices $n_i$. 
This sum can be done using the approximation
\be\label{approx1}
\sum_{n_1=k+1}^{l}\sum_{n_2=n_1+1}^{l}\dots\sum_{n_r=n_{r-1}+1}^{l}
p_{n_1}^2p_{n_2}^2 \dots p_{n_r}^2
\simeq \frac{1}{r!}\left(\sum_{m=k+1}^{l}p^2_m\right)^r,
\ee
which is actually exact in the continuum limit (i.e. when sums are replaced by integrals).
Therefore, inserting Eq. (\ref{result1}) in Eq. (\ref{corfun1}) and using Eq. (\ref{approx1}), we finally get
\bea
\langle N|\hat c^{\dag}_{k} \hat c_{l}|N\rangle 
& = & 2 \frac{N}{M}p_kp_l\sum_{r=0}^{N-1}(-2)^r\frac{(N-1)\dots(N-r)}{r!}
\left(\frac{2}{M}\sum_{m=k+1}^{l}p^2_m\right)^r \nonumber \\
& = & 2 \frac{N}{M} p_kp_l\left[1-2\left(\frac{2}{M}\sum_{m=k+1}^{l}p^2_m\right)\right]^{N-1}.
\label{corfun2}
\eea
Using now $\langle \hat c^{\dag}_k \hat c_l\rangle=\delta\langle \hat \Psi^{\dag}(x) \hat \Psi(y)\rangle$,
and $\delta \sum_{m} \equiv \int dz$ with $z=\delta m$, 
we can take the continuum limit of  Eq. (\ref{corfun2}), obtaining (for $x<y$)
\bea
\langle\hat\Psi^{\dag}(x)\hat \Psi(y)\rangle
& = & 2 \frac{N}{M\delta} p_k p_l
\left[1-2\left(\frac{2}{M\delta} \sum_{m=k+1}^{l}\delta \, p^2_m\right)\right]^{N-1} \nonumber \\
& = & 2\frac{N}{L} \sin\left(\frac{\pi}{L}x\right)\sin\left(\frac{\pi}{L}y\right)
\left[1-2\int_{x}^{y}dz \, \frac{2}{L}\sin^2\left(\frac{\pi}{L} z \right)\right]^{N-1},
\eea
which coincides with Eq. (\ref{PsiPsi_0_v2}).

%%%%%%%%%%%%%%% GENERAL TRAPPING POTENTIALS %%%%%%%%%%%%%%%%

\section{Generic confining potential}
\label{appB}

The initial correlation function given in  Eq. (\ref{PsiPsi_0_v2}) has two special limits which are not exclusive features of the hard-wall 
confining potential but are  valid for a generic potential (as long as the initial interaction is set to $c=0$), 
as we are going to show in this appendix. 
 
Let us consider a  generic trapping potential centred in $x=0$. 
The potential introduces a typical length-scale $\ell>0$ \cite{cv-11} (for example, in the presence of a harmonic confinement 
$V(x)=\omega^2 x^2 /2$ the typical length is $\ell \simeq 1/\sqrt{\omega}$), such that
the eigenfunctions  vanish for $|x|\gg \ell$.
The orthonormal one-particle eigenfunctions can be written as 
 \be
 \phi_p(x)=\frac{1}{\sqrt{\ell}}\psi_p\left(\frac{x}{\ell}\right),\quad p=0,1,2,\dots ,
\ee
where $\psi_p(z)$ are the normalised eigenfunctions for $\ell=1$.

We again consider as initial state the BEC constructed by placing $N$ particles in the lower energy-level with
wave-function $\phi_0(x)$.
Following the same logic which led us to Eq. (\ref{PsiPsi_0_v2}), we  find
the general form for the fermionic correlation function
\bea\label{PsiPsi_trap}
\langle\hat\Psi^{\dag}(x)\hat\Psi(y)\rangle 
& = & N \phi^{*}_0(x)\phi_0(y)\left[1-2 \left| \int_{x}^{y} dz \, |\phi_0(z)|^2 \right| \right]^{N-1}\nonumber\\
& = & (N/\ell) \psi^{*}_0(x/\ell)\psi_0(y/\ell)\left[1-(2/\ell) \left| \int_{x}^{y} dz \, |\psi_0(z/\ell)|^2 \right| \right]^{N-1},
\eea
which is valid for any finite $N$ and $\ell$.
Let us introduce the integral function  
\be
F(x)\equiv \int_{0}^{x} dz \, |\phi_0(z)|^2 =  \frac{1}{\ell}\int_{0}^{x} dz \, |\psi_0(z/\ell)|^2 
=  \int_{0}^{x/\ell} d\tilde z \, |\psi_0(\tilde z)|^2 \equiv \tilde F (x/\ell),
\ee
which satisfies $|\tilde F(y/\ell) - \tilde F(x/\ell)| \leq 1 $. 
$F(x)$ is a bounded monotonic even function, therefore $F(x)=F(y)$ implies $|x|=|y|$. 
We consider the thermodynamic limit $N\to\infty$, $\ell\to\infty$ with $N/\ell=n$, and with the additional
constraint that the rescaled variables $x/\ell$ and $y/\ell$  are kept finite.
Now, since $\psi_0(z)$ is a one-particle ground-state function with no nodes in its domain (apart from the boundaries if the domain is finite),
the limit 
\be
\lim_{N\to\infty} N\left[1-2 \left| \tilde F(y/\ell) - \tilde F(x/\ell) \right| \right]^{N-1}= \ell \frac{\delta(x-y)}{|\psi_0(x/\ell)|^2},
\ee
 leads to 
\be
\langle\hat\Psi^{\dag}(x)\hat\Psi(y)\rangle = \delta(x-y), \quad {\rm for} \quad N\to\infty.
\ee

However, this result does not correspond to the correct scaling regime and indeed applies only to the case of a very tight confining potential 
with an  extremely localised two-point fermionic function, in which the details of the trapping are lost.  
The fermionic mode occupation $\langle \hat n_{q} \rangle$ corresponding to this two-point correlator is $\langle \hat n_{q} \rangle=1$ 
which clearly is not physical.

In order to circumvent this problem, we could think of taking the TDL  by considering the variables $x$ and $y$ finite. 
In this case Eq. (\ref{PsiPsi_trap}) can be rewritten as
\be
\langle\hat\Psi^{\dag}(x)\hat\Psi(y)\rangle 
= n\psi^{*}_0(x/\ell)\psi_0(y/\ell)
\left[1- 2 \left| \tilde F(y/\ell) - \tilde F(x/\ell) \right|  \right]^{N-1},
\ee
which, for $\ell \gg 1$, with $x$ and $y$ fixed, can be expanded around $x,y\sim 0$ as
\be
\langle\hat\Psi^{\dag}(x)\hat\Psi(y)\rangle 
= n|\psi_0(0)|^2
\left[1- \frac{2 n}{N} |\psi_0(0)|^2 | x-y | \right]^{N-1}.
\ee
and finally, using $\lim_{N\rightarrow\infty}\left(1 + z/N\right)^N = {\rm e}^{z}$, one obtains
\be\label{PsiPsi_trap2}
\langle\hat\Psi^{\dag}(x)\hat\Psi(y)\rangle 
=  n|\psi_0(0)|^2 {\rm e}^{-2 n|\psi_0(0)|^2 |x-y|},
\ee
which coincides with the result for PBC \cite{kcc14} (when $|\psi_0(0)|^2 = 1$). 
This result is easily understood:  for $\ell \gg 1$ and $x$ and $y$ finite, 
the system retains only information about the value of the initial density in the middle of the trap, i.e.  Eq. (\ref{PsiPsi_trap2}) 
is equivalent to consider a translational invariant case with homogeneous initial density equals to $|\psi_0(0)|^2$ thence
losing completely the effect of the trap. 

Thus we conclude that there are no shortcuts in this problem and,  in order to properly retain the confinement effects, 
the correct way to proceed is to keep  $N$ and $\ell$ finite
(which in our specific case corresponds to the size $L$) for the calculation
of $\langle \hat n_{q} \rangle$ and $\langle \hat\eta^{\dag}_p \hat\eta_q \rangle$, and only
afterwards take the thermodynamic limit.

%%%%%%%%%%%%%%% APPROXIMATIONS %%%%%%%%%%%%%%%%

\section{Technical details for the evaluation of the time-dependent correlation function}\label{appC}

In this appendix we show the details of the calculations needed to derive   
Eq. (\ref{Cxt_2}) from Eq. (\ref{Cxt_1}).

From well known trigonometric identities, we can rewrite the term
\be
\sin\left[\frac{p\pi}{2L}(w+z)\right]\sin\left[\frac{q\pi}{2L}(w-z)\right],
\ee
as
\bea\label{trig_ident}
\frac{1}{2}\cos\left[\frac{\pi(p+q)z}{2L}\right] \cos\left[\frac{\pi(p-q)w}{2L}\right] & - & 
\frac{1}{2}\cos\left[\frac{\pi(p-q)z}{2L}\right] \cos\left[\frac{\pi(p+q)w}{2L}\right]+ \nonumber \\
\frac{1}{2}\sin\left[\frac{\pi(p+q)z}{2L}\right] \sin\left[\frac{\pi(p-q)w}{2L}\right] & - & 
\frac{1}{2} \sin\left[\frac{\pi(p-q)z}{2L}\right] \sin\left[\frac{\pi(p+q)w}{2L}\right].
\eea
In the thermodynamic limit the only relevant contribution to the time-dependent
correlation function comes from the first term of Eq. (\ref{trig_ident}). 
All the other terms are either identically vanishing or  introduce finite-size corrections which disappear in the TDL. 
Indeed, since we are working in the regime $p+q\gg1$, $w = x+y  \sim O(L)$ and $z=x-y\sim O(1)$, using the rescaled 
variables $\tilde w = w/L,\; \tilde r = (p+q)/(2L), \; l=(p-q)/2$, we have
\be\label{trig_ident_2}
\frac{1}{2}\cos(\pi \tilde r z) \cos(\pi l \tilde w) - 
\frac{1}{2}\cos\left(\frac{\pi l}{L} z\right) \cos( \pi L \tilde r \tilde w)+
\frac{1}{2}\sin(\pi \tilde r z)\sin(\pi l \tilde w) - 
\frac{1}{2} \sin\left(\frac{\pi l}{L} z\right) \sin(\pi L \tilde r \tilde w).
\ee
The third term  in Eq. (\ref{trig_ident_2}) does not contribute to the evaluation of the 
time-dependent correlation function since it is an odd function of $l$ and the sum over 
$l$ in Eq. (\ref{Cxt_2}) is symmetric around zero.
Moreover, as $L\to\infty$ the last term vanishes.
Therefore, the only terms which survive are
\be\label{trig_ident_3}
\frac{1}{2}\cos(\pi \tilde r z) \cos(\pi l \tilde w)-\frac{1}{2}\cos( \pi L \tilde r \tilde w).
\ee
The first term in Eq. (\ref{trig_ident_3}) is exactly the term that was considered in the main text
and which leads to the correct  time-dependent correlation function. 
The second one, instead, introduces only finite-size corrections; indeed, following the same reasoning as in
Sec. \ref{sec4}, it is straightforward to show that it corresponds to ($\tilde t = t/L$)
\be
-\frac{1}{2 \pi} \int_{-\infty}^{\infty} d\tilde r \sum_{l=-\infty}^{\infty} \int_{0}^{\pi} du \,
\cos( \pi L \tilde r \tilde w)
\frac{\cos[l(u+\pi)]}{1+\left[\frac{\tilde r\pi/(2n)}{1+\cos(u)}\right]^{2} }
{\rm e}^{i 4 \pi^2 \tilde r l \tilde t} \sim \exp(-L),
\ee
thus vanishing in the thermodynamic limit.

%%%%%%%%%%%%%%%%%%%%%%%%%%%%%%%%%%%%%%%%%%%%%%%%%%


\begin{thebibliography}{99}

\bibitem{uc}
M.~Greiner, O.~Mandel, T.~W.~H\"ansch, and I.~Bloch, Nature {\bf 419} 51 (2002).

\bibitem{kww-06}
T. Kinoshita, T. Wenger,  D. S. Weiss, %A quantum Newton's cradle,
 Nature {\bf 440}, 900 (2006).


\bibitem{tetal-11}
S. Trotzky Y.-A. Chen, A. Flesch, I. P. McCulloch, U. Schollw\"ock, J. Eisert, and I. Bloch, 
Nature Phys. {\bf 8}, 325 (2012). 

\bibitem{cetal-12}
M. Cheneau, P. Barmettler, D. Poletti, M. Endres, P. Schauss, T. Fukuhara, C. Gross, I. Bloch, C. Kollath, and S. Kuhr,
%Light-cone-like spreading of correlations in a quantum many-body system,
Nature {\bf 481}, 484 (2012).

\bibitem{getal-11}
M. Gring, M. Kuhnert, T. Langen, T. Kitagawa, B. Rauer, M. Schreitl, I. Mazets, D. A. Smith, E. Demler, and J. Schmiedmayer,
%Relaxation Dynamics and Pre-thermalization in an Isolated Quantum System,
Science {\bf 337}, 1318 (2012).

\bibitem{shr-12}
U. Schneider, L. Hackerm\"uller, J. P. Ronzheimer, S. Will, S. Braun, T. Best, I. Bloch, E. Demler, S. Mandt, D. Rasch, and A. Rosch,
Nature Phys. {\bf 8}, 213 (2012).


\bibitem{rsb-13}
J. P. Ronzheimer, M. Schreiber, S. Braun, S. S. Hodgman, S. Langer, I. P. McCulloch, F. Heidrich-Meisner, I. Bloch, and U. Schneider,
Phys. Rev. Lett. {\bf 110}, 205301 (2013).


\bibitem{revq}
A. Polkovnikov, K. Sengupta, A. Silva, and M. Vengalattore, 
Rev. Mod. Phys. {\bf 83}, 863 (2011).

%%%%%%%%%%%%

\bibitem{gg} 
M. Rigol, V. Dunjko, V. Yurovsky,  and M. Olshanii, Phys. Rev. Lett. {\bf 98}, 50405 (2007).


\bibitem{rs-12}
M. Rigol and M. Srednicki, Phys. Rev. Lett. {\bf 108}, 110601 (2012).


\bibitem{ce-13}
J.-S. Caux and F. H. L. Essler, Phys. Rev. Lett. {\bf 110}, 257203 (2013).


\bibitem{nonint} J. M. Deutsch, Phys. Rev. A {\bf 43}, 2046 (1991);\\
  M. Srednicki, Phys. Rev. E {\bf 50}, 888 (1994).

\bibitem{rdo-08}
M. Rigol, V. Dunjko, and M. Olshanii, Nature {\bf 452}, 854 (2008);\\
M. Rigol, Phys. Rev. Lett. {\bf 103}, 100403 (2009); \\
M. Rigol, Phys. Rev. A {\bf 80}, 053607 (2009).

\bibitem{tvar}
G. Biroli, C. Kollath, and A. Laeuchli, Phys. Rev. Lett. {\bf 105}, 250401 (2010);\\
M. C. Banuls, J. I. Cirac, and M. B. Hastings, Phys. Rev. Lett. {\bf 106}, 050405 (2011);\\
M. Rigol and M. Fitzpatrick, Phys. Rev. A {\bf 84}, 033640 (2011);\\
K. He and M. Rigol, Phys. Rev. A {\bf 85}, 063609 (2012);\\
G. P. Brandino, A. De Luca, R. M. Konik, and G. Mussardo, Phys. Rev. B {\bf 85}, 214435 (2012);\\
J. Sirker, N.P. Konstantinidis, and N. Sedlmayr, Phys. Rev. A {\bf 89}, 042104 (2014).

%%%%%%

\bibitem{LiebPR130} E. H. Lieb and W. Liniger, Phys. Rev. {\bf 130}, 1605 (1963);  \\
E. H. Lieb, Phys. Rev. {\bf 130}, 1616 (1963).




\bibitem{grd-10}
V. Gritsev, T. Rostunov, and E. Demler, J. Stat. Mech. (2010) P05012.

\bibitem{fle-10}
D. Muth, B. Schmidt, and M. Fleischhauer, New J. Phys. {\bf 12}, 083065 (2010); \\
D. Muth and M. Fleischhauer, Phys. Rev. Lett. {\bf 105}, 150403 (2010).

\bibitem{mc-12}
J. Mossel and J.-S. Caux, New J. Phys. {\bf 14},  075006 (2012).

\bibitem{ksc-13}
M. Kormos, A. Shashi, Y.-Z. Chou, J.-S. Caux, and A. Imambekov,
Phys. Rev. B {\bf 88}, 205131 (2013).  

\bibitem{nm-13}
S. S. Natu and E. J. Mueller, 	Phys. Rev. A {\bf 87}, 053607 (2013).

\bibitem{kcc14}
M. Kormos, M. Collura and P. Calabrese, Phys. Rev. A {\bf 89}, 013609 (2014).

\bibitem{nwbc-13}
J. De Nardis, B. Wouters, M. Brockmann, and J.-S. Caux, Phys. Rev. A {\bf 89}, 033601 (2014).

\bibitem{ds-13}
P. Deuar and M. Stobinska, arXiv:1310.1301.

\bibitem{ckc-14}
M. Collura, M. Kormos,  and P. Calabrese, J. Stat. Mech. P01009 (2014).

\bibitem{b-14}
M. Brockmann, J. Stat. Mech. (2014) P05006;\\
M. Brockmann, J. De Nardis, B. Wouters, and J.-S. Caux,  arXiv:1403.7469.

\bibitem{cd-14}
P. Calabrese and P. Le Doussal, J. Stat. Mech. (2014) P05004.


\bibitem{mg-05}
A. Minguzzi and D.M. Gangardt, Phys. Rev. Lett. {\bf 94}, 240404 (2005).

\bibitem{cro}
H. Buljan, R. Pezer, and T. Gasenzer, Phys. Rev. Lett. {\bf 100}, 080406 (2008).

\bibitem{ck-12}
J.-S. Caux and R. M. Konik, Phys. Rev. Lett. {\bf 109}, 175301 (2012).

\bibitem{a-12}
D. Iyer and N. Andrei, Phys. Rev. Lett. {\bf 109}, 115304 (2012);\\
D. Iyer, H. Guan, and N. Andrei, Phys. Rev. A {\bf 87}, 053628 (2013).

\bibitem{v-12}
E. Vicari, Phys. Rev. A {\bf 85},   062324 (2012).

\bibitem{csc13}
M. Collura, S. Sotiriadis, and P. Calabrese, Phys. Rev. Lett. {\bf 110}, 245301 (2013);\\
M. Collura, S. Sotiriadis, and P. Calabrese, J. Stat. Mech. (2013) P09025.

\bibitem{m-13}
G. Mussardo, Phys. Rev. Lett. {\bf 111}, 100401 (2013).

\bibitem{gn-14b}
G. Goldstein and N. Andrei, arXiv:1406.4902.


\bibitem{hwint}
M. Gaudin, Phys. Rev. A {\bf 4}, 386 (1971);\\
M. Gaudin, La fonction dÕonde de Bethe (1983 Paris: Masson);\\
M.T. Batchelor, X.W. Guan, N. Oelkers, and C. Lee, J. Phys. A {\bf 38}, 7787 (2005).


\bibitem{TG} L. Tonks, Phys. Rev. {\bf 50}, 955 (1936); \\
 M. Girardeau, J. Math. Phys. {\bf 1}, 516 (1960).




\bibitem{cdeo-08}
M. Cramer, C. M. Dawson, J. Eisert, and T. J. Osborne,  Phys. Rev. Lett. {\bf 100}, 030602 (2008);\\
M. Cramer and J. Eisert, New J. Phys. 12, 055020 (2010).

\bibitem{bs-08}
T. Barthel and U. Schollw\"ock,  
Phys. Rev. Lett. {\bf 100}, 100601 (2008).

\bibitem{CEF}
P. Calabrese, F. H. L. Essler, and M. Fagotti, Phys. Rev. Lett. {\bf 106}, 227203 (2011).

\bibitem{CEFII}
P. Calabrese, F. H. L. Essler, and M. Fagotti, 
J. Stat. Mech. (2012) P07022.





%==========

\bibitem{f-14}
M. Fagotti, J. Stat. Mech. (2014) P03016.

\bibitem{noGGE1}
B. Wouters, M. Brockmann, J. De Nardis, D. Fioretto and J.-S. Caux, arXiv:1405.0172.

\bibitem{noGGE2}
 B. Pozsgay, M. Mestyan, M. A. Werner, M. Kormos, G. Zarand and G. Takacs, arXiv:1405.2843.

\bibitem{noGGEp}
M. Mierzejewski, P. Prelovsek, and T. Prosen, arXiv:1405.2557.

\bibitem{noGGE3}
G. Goldstein and N. Andrei, arXiv:1405.4224.

\bibitem{noGGE4}
 B. Pozsgay, arXiv:1406.4613
 
\bibitem{fe-13b}
M.  Fagotti and  F. H. L. Essler, J. Stat. Mech. (2013) P07012.

\bibitem{fcce-13}
M. Fagotti, M. Collura, F. H. L. Essler, and P. Calabrese, Phys. Rev. B {\bf 89}, 125101 (2014).

%%%%%%%%%%%%%%%%%%%


\bibitem{fe-13}
M.  Fagotti and  F. H. L. Essler, Phys. Rev. B {\bf 87}, 245107 (2013).


\bibitem{sc-14}
S. Sotiriadis and P. Calabrese, arXiv:1403.7431.


\bibitem{eef-12}
F. H. L. Essler, S. Evangelisti, and M. Fagotti, Phys. Rev. Lett. {\bf 109}, 247206 (2012).


\bibitem{fredholm}
A. Jerri, \textit{Introduction to Integral Equations with Applications}, John Wiley \& Sons (1999);\\
J. Feinberg, J. Phys. A: Math. Gen. {\bf 37}, 6299 (2004).

\bibitem{adi}
A. Imambekov, I. E. Mazets, D. S. Petrov, V. Gritsev, S. Manz, S.Hofferberth, T. Schumm, E. Demler, and J. Schmiedmayer,
Phys. Rev. A {\bf 80}, 033604 (2009).

\bibitem{ck14}
M. Collura and D. Karevski, Phys. Rev. B. {\bf 89}, 214308 (2014).

\bibitem{cc-06}
P. Calabrese and  J. Cardy,  Phys. Rev. Lett. {\bf 96}, 136801 (2006); \\
P. Calabrese and  J. Cardy,  J. Stat. Mech. P06008  (2007); \\
P. Calabrese and  J. Cardy, J. Stat. Mech. P04010 (2005).

\bibitem{fm-10}
D. Fioretto and G. Mussardo, New J. Phys. {\bf 12}, 055015 (2010).

\bibitem{sfm-12}
S. Sotiriadis, D. Fioretto, and G. Mussardo, J. Stat. Mech. (2012) P02017.


\bibitem{CEFI}
P. Calabrese, F. H. L. Essler, and M. Fagotti, J. Stat. Mech. (2012) P07016.


\bibitem{ir-10}
F. Igloi and H. Rieger, Phys. Rev. Lett. {\bf 106}, 035701 (2011);\\
H. Rieger and F. Igloi, Phys. Rev. B {\bf 84}, 165117 (2011).

\bibitem{cv-11}
M. Campostrini and E. Vicari, Phys. Rev. A {\bf 81} 023606 (2010);\\
M. Campostrini and E. Vicari, Phys. Rev. A {\bf 81} 063614 (2010).

 
















\end{thebibliography}
\end{document}